# Local-time Dependence of Chemical Species in the Venusian Mesosphere

Wencheng D. Shao[*,1], Xi Zhang[1], João Mendonça[2], Thérèse Encrenaz[3]

**Abstract**

Observed chemical species in the Venusian mesosphere show local-time variabilities. $SO_2$ at the cloud top exhibits two local maxima over local time, $H_2O$ at the cloud top is uniformly distributed, and CO in the upper atmosphere shows a statistical difference between the two terminators. In this study, we investigated these local-time variabilities using a three-dimensional (3D) general circulation model (GCM) in combination with a two-dimensional (2D) chemical transport model (CTM). Our simulation results agree with the observed local-time patterns of $SO_2$, $H_2O$, and CO. The two-maximum pattern of $SO_2$ at the cloud top is caused by the superposition of the semidiurnal thermal tide and the retrograde superrotating zonal (RSZ) flow. $SO_2$ above 85 km shows a large day-night difference resulting from both photochemistry and the sub-solar to anti-solar (SS-AS) circulation. The transition from the RSZ flows to SS-AS circulation can explain the CO difference between two terminators and the displacement of the CO local-time maximum with respect to the anti-solar point. $H_2O$ is long-lived and exhibits very uniform distribution over space. We also present the local-time variations of HCl, ClO, OCS and SO simulated by our model and compare to the sparse observations of these species. This study highlights the importance of multidimensional CTMs for understanding the interaction between chemistry and dynamics in the Venusian mesosphere.

[*]Corresponding author: Wencheng D. Shao (wshao7@ucsc.edu)
[1]Department of Earth and Planetary Sciences, University of California, Santa Cruz, CA 95064, USA.
[2]National Space Institute, Technical University of Denmark, Elektrovej, DK-2800 Kgs. Lyngby, Denmark
[3]LESIA, Observatoire de Paris, PSL University, CNRS, Sorbonne University, University Sorbonne Paris City, 92195 Meudon, France

# 1. MOTIVATION

The Venusian mesosphere (~70-100 km) is characterized by complex photochemistry (e.g., Yung & DeMore 1982; Mills 1998; Zhang et al. 2012; Krasnopolsky 2012). Species like $SO_2$ and CO — fundamental components of the photochemical network — have shown significant spatial and temporal variabilities. For example, Venus Express detected that the $SO_2$ mixing ratio at 70-80 km varies by orders of magnitude over time and space (Vandaele et al. 2017a, 2017b). Ground-based observations by TEXES/IRTF (Texas Echelon Cross Echelle Spectrograph/Infrared Telescope Facility) show that $SO_2$ around the cloud top exhibits plumes and patchy features over the Venus disk (Encrenaz et al. 2012, 2013, 2016, 2019, 2020). Venus Express also observed that CO has strong short-term variabilities up to one order of magnitude (Vandaele et al. 2016).

Some chemical species in the Venusian mesosphere show strong local-time variabilities. Encrenaz et al. (2020) extracted the local-time dependence of $SO_2$ at ~64 km from TEXES/IRTF and found that the $SO_2$ mixing ratio generally exhibits two local maxima around the morning and evening terminators, respectively. SPICAV (Spectroscopy for the investigation of the characteristics of the atmosphere of Venus) on board Venus Express also observed a similar $SO_2$ local-time pattern at the cloud top on the dayside (Vandaele et al. 2017b; Encrenaz et al. 2019; Marcq et al. 2020). Sandor et al. (2010) used microwave spectra to obtain day-night differences of $SO_2$ and SO at 70-100 km. Despite the scarcity of the data, $SO_2$ appears more abundant at night than that during the day, while SO is likely to have a reversed day-night difference. Belyaev et al. (2017) observed that midnight $SO_2$ abundance appears 3-4 times higher than at the terminators around 95 km through SPICAV occultations. For CO, Clancy and Muhleman (1985) observed a day-night difference from microwave measurements. They found that the CO bulge (i.e., the local maximum of CO mixing ratio) shifts from midnight to the morning as altitude decreases from above ~95 km to 80-90 km. Clancy et al. (2003) showed similar CO patterns in subsequent microwave observations.

Vandaele et al. (2016) summarized the CO data observed by SOIR (Solar Occultation in the InfraRed) on board Venus Express and found a statistical difference between the morning and evening terminators, and the difference also depends on altitude. Compared to $SO_2$ and CO, $H_2O$ seems to vary insignificantly with local time. Encrenaz et al. (2012, 2013, 2016, 2019, 2020) using TEXES observed that the $H_2O$ mixing ratio at ~64 km, obtained from the HDO spectra, distributes uniformly over the Venus disk. Chamberlain et al. (2020) showed that the $H_2O$ profiles above 80 km observed by SOIR do not exhibit dependence on terminators. Sandor and Clancy (2012, 2017) using JCMT (James Clerk Maxwell Telescope) observed that HCl mixing ratio above 85 km exhibts no evident day-night difference. Krasnopolsky (2010) using the CSHELL spectrograph at NASA IRTF observed that the morning OCS can be more abundant than the afternoon OCS.

The origin of these local-time variabilities has not been thoroughly investigated but likely relates to atmospheric chemistry and dynamics. In the Venusian mesosphere occurs intense photochemistry (e.g., Zhang et al. 2012), in which the dependence of solar irradiance on local time affects the local distribution of chemical species. On Venus, the cloud region (~47-70 km) is characterized by a retrograde superrotating zonal (RSZ) flow (e.g., Sánchez-Lavega et al. 2008; Lebonnois et al. 2010; Mendonça & Read 2016; Mendonça & Buchhave 2020). In the thermosphere (>110 km), strong evidence shows a sub-solar to anti-solar (SS-AS) circulation pattern (e.g., Bougher et al. 2006). The upper mesosphere (90-110 km) might be a region where SS-AS circulation is superimposed on the RSZ flow (e.g., Lellouch et al. 1994). Besides, thermal tides excited by the solar heating also strongly perturb the temperature and winds in the mesosphere (e.g., Taylor et al. 1980; Limaye 2007; Fukuya et al. 2021). These dynamical flow patterns transport chemical species and modulate their local-time variabilities.

A few theoretical studies have investigated the local-time variabilities of chemical species. Jessup et al. (2015) studied spatial variations of $SO_2$ and SO observed by HST/STIS (Hubble Space Telescope Imaging Spectrograph). They showed through one-dimensional (1D) photochemical models that solar zenith angle could significantly affect the $SO_2$ variability. Gilli et al. (2017) presented CO and O density profiles in the upper atmosphere at different local times using a three-dimensional (3D) general circulation model (GCM) coupling chemistry and dynamics (Stolzenbach et al. 2015; Stolzenbach 2016). Their results indicate the importance of the SS-AS circulation on the CO and O distributions. Navarro et al. (2021) and Gilli et al. (2021) used an improved GCM to study CO's spatial variabilities in the upper atmosphere. Their simulated CO pattern shows a CO bulge shift toward the morning by 2-3 hours in the mesosphere, attributed to a weak westward retrograde wind. However, a dedicated study of the local-time variability of $SO_2$ is still lacking, and mechanisms controlling CO's local-time distributions need further investigation.

As a preliminary step towards fully understanding the spatial and temporal variabilities of chemical species in the Venusian atmosphere, in this study we investigate the local-time dependence of multiple chemical species including $SO_2$ and CO using a 3D dynamical model in combination with a 2D (longitude-pressure) chemical model with the state-of-the-art full photochemical network on Venus. Our simulated local-time distributions of species like $SO_2$, CO, and $H_2O$ show agreement with observations. We explore underlying mechanisms determining species' local-time distributions and find that the relative importance of dynamics and chemistry depends on altitude and species. Our study indicates that the local-time distributions of $SO_2$ and CO can constrain important dynamical patterns in the Venusian atmosphere.

This paper is structured as follows. Section 2 and Appendix A provide technical details of our models. In Section 3, we present simulations from our nominal case and study the local-time dependence of $SO_2$, CO and other species. In Section 4, we do sensitivity tests and discuss the influences of our models' parameters and resolution on our results. Finally, we conclude our results and discuss future work in Section 5.

## 2. METHODOLOGY

We use a 3D GCM in combination with a 2D chemical-transport model (CTM) to study the local-time variabilities of chemical species in the Venusian mesosphere. We adopt this combination method because fully coupling chemistry with dynamics in the GCM or utilizing a 3D CTM with a full chemical network is computationally expensive. A 2D (longitude-pressure) chemical model is sufficient to study the chemical species' local-time variabilities that we focused on in this work. For example, combining a 3D GCM with a 2D CTM has been used to study chemical species in Earth's atmosphere (e.g., Smyshlyaev et al. 1998).

We adopt the OASIS GCM, a novel and flexible 3D planetary model (Mendonça & Buchhave 2020). OASIS is a dedicated model that incorporates multiple self-consistent modules. For our Venus dynamical simulations, we use the non-hydrostatic dynamical core coupled with physics modules that represent a basalt soil/surface, convective adjustments, and the radiative processes from the gas and clouds (a non-grey scheme with multiple-scattering). The simulated atmosphere extends from the surface to 100 km, with a horizontal resolution of 2 degrees and a vertical resolution of ~2 km. The model was integrated for 25000 Earth days (~214 Venus solar days; one Venus solar day is ~117 Earth days) with a time-step of 50 seconds. The model and bulk planet parameters (e.g., specific heat, gravity and mean radius) are the same as the ones used in Mendonça and Buchhave (2020, see their Table 2). One of the main weaknesses of current Venus GCMs is the poor representation of the circulation in the deep atmosphere, which is also poorly constrained by observational data (refer to Mendonça & Read 2016 for more details). To represent a deep circulation in our 3D simulations closer to the observations, we applied a Newtonian relaxation method to force the zonal winds in the deep atmosphere towards the observed values. The forcing acts only at 44 km altitude, which is below the cloud region and the region explored in this study. The equilibrium winds were constructed assuming the atmosphere at 44 km rotating as a solid body with a maximum velocity of 50 m/s at the equator based on the estimated observed values from Kerzhanovich & Limaye (1985). For the Newtonian relaxation timescale, we have assumed a value of 2000 Earth days (~17 Venus solar days), which is close to the radiative timescale at 44 km (Pollack & Young 1975). At 44 km, the temperature difference between the day- and night-side of the planet is small (less than 10K) because the radiative timescale is much longer than the dynamical timescale. The value chosen for the relaxation timescale ensures a good model performance and low impact in the wave activity in the lower atmosphere. Our converged simulations were further integrated to 5000 Earth days (~43 Venus solar days) to produce the temperature and wind fields for the CTM input.

We do not directly couple the GCM and CTM in the sense that the simulated gas distributions in the CTM are not used as the GCM input. As described in Mendonca and Buchhave (2020), the 3D GCM itself only uses simple representations of the clouds and chemistry. The cloud structure remains constant with time, and three different cloud particle size modes (Knollenberg & Hunten 1980; Crisp 1986) are used. The GCM considers four main chemical species in the atmosphere: 96.5% of $CO_2$ in mole, ~3.5% of $N_2$, 50 ppm of $H_2O$, and 100 ppm of $SO_2$. Their volume mixing ratios are assumed to be well-mixed in the GCM and not meant to be exactly equal to the values observed but to capture the main bulk conditions of the Venusian atmosphere.

The 2D CTM is generalized from the 1D state-of-the-art Caltech/JPL kinetics model (Yung & DeMore 1982; Mills 1998; Mills & Allen 2007; Zhang et al. 2010, 2012; Bierson & Zhang 2020; Shao et al. 2020). This model resolves complex chemistry for carbon, oxygen, hydrogen, nitrogen, sulfur, and chlorine species. This model includes 52 chemical species and over 400 reactions (refer to Zhang et al. 2012). We generalized this 1D model to 2D and included the advection for each chemical tracer in the longitude-(log-)pressure coordinate plane. See

Appendix A for the derivation of the meridionally-mean continuity equation. The meridionally-mean advection is constructed from the output of the 3D GCM. We implemented the flux-limiting Prather scheme (Prather 1986; Shia et al. 1989, 1990) to calculate the advection of species in the 2D continuity equation. This scheme has several advantages, including the conservation of chemical species, maintenance of positive concentration, and stability for large time steps. The chemical model incorporating this scheme has been applied to the Earth's atmosphere to study variabilities of chemical species like ozone (Jiang et al. 2004). We have also implemented a parallel computing technique using Message Passing Interface (MPI) in our 2D CTM to improve simulation efficiency.

In our 2D CTM, photon density reaching the top of atmosphere (TOA) is set as equal to $\pi/4$ times the equatorial value on Venus, so as to represent the meridionally-mean value considering the latitudinal dependence of solar zenith angle. The solar zenith angle in our 2D CTM varies with longitude, and the solar zenith angle at each longitude also changes with time. Our 2D CTM has a vertical resolution of ~2 km and a horizontal resolution of 12 degrees. The altitude range is ~58-100 km. The time step is set as 10 minutes. In Section 4, we show that increasing horizontal resolution does not change the simulated local-time variabilities of chemical species.

The 3D distributions of temperature and wind patterns from GCM simulations in the last ~4 Venus solar days are first averaged meridionally. To match the spatial and temporal grids in our 2D CTM, we then smooth and interpolate the GCM data to obtain temperature and wind fields in one-hour resolution. Finally, we average the fields temporally to obtain the diurnally-varying one-Venus-day (~117 Earth days) fields and repeatedly input them into the 2D CTM. Table 1 lists boundary conditions for several important species. For other species, zero flux at the upper boundary and maximum deposition velocity at the lower boundary (58 km) are applied. We apply the same lower and upper boundary conditions to all longitudes. The unknown sulfur reservoir in the upper atmosphere (e.g., Zhang et al. 2010, 2012; Vandaele et al. 2017a) is represented by a downward $S_8$ flux at the upper boundary in our model, as used in Bierson and Zhang (2020). The specified flux at the upper boundary (e.g., Table 1) is separate from the advective flux and is used to provide extra sources outside the domain (e.g., the $S_8$ flux). We calculate the advective flux above (below) the upper (lower) boundary by setting a ghost box with species' mixing ratios the same as those at the boundary. In the zonal direction, a periodic boundary condition (i.e., species abundances at 0 and 360 degrees are equal) is adopted.

In this study, we treat the sub-grid diffusivity parameters $K_{xx}$, $K_{xz}$, $K_{zx}$, and $K_{zz}$ in the meridionally-mean continuity equation (see Appendix A) as free parameters. For simplicity, we assume zero $K_{xz}$ and $K_{zx}$. The meridionally-mean zonal wind is usually larger than the eddy wind. For example, the meridionally-mean zonal wind is ~100 m/s at ~60 km, while the eddy wind is ~10-20 m/s at ~60 km at the equatorial region in the GCM output. If we assume the sub-grid eddy length scale is 10-100 km (the horizontal grid size is about 200 km around the equator in the GCM), the horizontal diffusivity $K_{xx}$ is about $10^9 - 10^{10}\ cm^2 s^{-1}$. Here we use $K_{xx} = 10^9\ cm^2 s^{-1}$ to represent the horizontal transport by eddies. The $K_{zz}$ vertical profile in our CTM is the same as the 1D $K_{zz}$ profile in Zhang et al. (2012) and is applied to all longitudes. In Section 4, we will explore the sensitivity of our results to these parameters.

### 3. LOCAL-TIME DEPENDENCE OF CHEMICAL SPECIES

In this section, we discuss the local-time dependence of $SO_2$, CO, $H_2O$, HCl, ClO, OCS and SO. These chemical species' distributions are averaged over the last 5 Venus solar days from our simulations. To better understand the interaction between atmospheric dynamics and chemistry and the influence on local-time variability, we also estimate the chemical loss timescale $t_{Chemical}$ and horizontal transport timescale $t_{Transport}$. The chemical loss timescale for each species is equal to the number density of the species divided by its total chemical loss rate. For $SO_2$, CO and SO, the chemical loss in the fast cycles (see Appendix B) is excluded. The horizontal transport timescale is estimated using the planetary radius divided by the zonal wind speed (e.g., Zhang & Showman 2018).

The 3D GCM results show that the diurnal cycle excites various harmonics of the thermal tides in the Venus atmosphere (Figure 1a-b). The thermal tides are stationary with respect to the sub-solar point. Observational (e.g., Pechmann & Ingersoll 1984; Zasova et al. 2007) and theoretical (e.g., Lebonnois et al. 2010; Mendonça & Read 2016) studies suggest that the semidiurnal component has the largest amplitude of the thermal tide harmonics in the upper cloud region. Our GCM successfully simulates this semidiurnal component in the upper cloud (58-70 km). In the vertical wind pattern shown in Figure 1b, the semidiurnal tide at 58-70 km induces an upwelling branch in the afternoon. A similar branch also appears at 0:00-6:00 but with a weaker amplitude. At 18:00-0:00 and 6:00-12:00, the semidiurnal tide in the upper cloud induces downwelling motions. The evening downwelling is stronger than the morning downwelling. Positive temperature anomalies are found around midnight and noon as a result of perturbations by the semidiurnal tide.

Above 85 km, the diurnal thermal tide dominates the wind and temperature local variations in our simulations (Figure 1a-b). As altitude increases, the phase of the semidiurnal tide shifts eastward. Above 85 km, two upwelling branches of the semidiurnal tide merge into a dayside upwelling branch, while the morning downwelling of the semidiurnal tide extends and becomes the nightside downwelling branch above 85 km. In this altitude region, the vertical wind field is mainly composed of the wavenumber-one diurnal component. Temperature distribution is also affected by the diurnal thermal tide above 85 km, with positive anomaly on the dayside. At 85-100 km, the SS-AS circulation is imposed on the RSZ flows, marking a transition region from RSZ flow to SS-AS circulation where the wind pattern is important for chemical tracer exchange between the lower and higher altitudes as well as that between the dayside and the nightside. The chemical tracers at the lower altitude are first transported upward by the upwelling branch on the dayside and then delivered to the nightside by horizontal day-night flows of the SS-AS circulation. On the nightside, the chemicals are recycled back to the lower region by the downwelling branch of the SS-AS circulation and transported to the dayside by the RSZ flows.

### 3.1 $SO_2$

Our simulation shows that the semidiurnal thermal tide is essential to explain the $SO_2$ local-time pattern in the upper cloud region (Figure 1). Below 80 km, the $SO_2$ mixing ratio decreases as altitude increases. Vertical mixing occurs when the $SO_2$-rich air is transported upward from a lower altitude, and the $SO_2$-poor air is transported downward from a higher altitude. In the upper cloud region (58-70 km), the two upwelling branches of the semidiurnal tide produce two local maxima in the $SO_2$ local-time distribution. The two maxima are shifted westward by the RSZ flow and located around two terminators. This local-time pattern is more clearly seen in Figure 2. Our simulations successfully reproduce the $SO_2$ observations by TEXES at ~64 km (Encrenaz et al. 2020). Note that SPICAV also observed a similar $SO_2$ local-time distribution at ~70 km on the dayside (Vandaele et al. 2017b; Encrenaz et al. 2019; Marcq et al. 2020). The TEXES data exhibit more complicated features, like a peak around 22 hour and another around 2 hour (Figure 2). The causes of these peaks are not well understood and might be associated with small-scale dynamics.

In the region above the clouds, photochemistry drives the $SO_2$ behavior on the dayside (Figure 1f). Above 85 km, $SO_2$ day-night difference becomes evident; $SO_2$ is less abundant on the dayside than on the nightside. Both photochemistry and dynamics drive this day-night difference. On the dayside, photolysis destroys $SO_2$; on the nightside, the descending branch of the SS-AS circulation brings $SO_2$-rich air downward because $SO_2$ mixing ratio generally increases as altitude increases above 85 km due to the assumed $S_8$ downward flux in our model. Sandor et al. (2010) implied an $SO_2$ day-night difference from microwave measurements at 70-100 km, despite the scarcity of their data. Belyaev et al. (2017) observed 150-200 ppb $SO_2$ at midnight versus 50 ppb $SO_2$ at terminators around 95 km from SPICAV occultations. Our simulations roughly agree with the observed $SO_2$ pattern around 95 km (Figure 10a).

### 3.2 CO

CO is a long-lived chemical species whose local-time distribution is determined mainly by dynamics (Figure 3). Figure 3b and 3c show that CO has a long chemical loss timescale above the clouds. Below 85 km, CO is well-mixed and almost exhibits no diurnal variations. Above 85 km, CO shows a day-night difference similar to $SO_2$ (Figure 3a). This difference is caused by the SS-AS circulation. The CO mixing ratio generally increases as altitude increases above 85 km because it is mainly produced by $CO_2$ photolysis in the upper atmosphere and transported downward. Above 85 km, the SS-AS circulation reduces CO on the dayside by mixing the CO-poor air upward from below. Although photochemistry produces more CO on the dayside, the SS-AS circulation regulates and dominates the CO local-time distribution by transporting the CO from dayside to the nightside. As a result, CO is accumulated on the nightside and appears more abundant than the dayside.

The CO maximum is located around midnight at ~95 km and shifts to dawn at ~85 km (also see Figure S3c-d). This pattern has been observed by microwave instruments (Clancy & Muhleman 1985; Clancy et al. 2003). The CO maximum is shifted westward due to zonal winds in the transition region, where SS-AS circulation transits to RSZ flow as altitude decreases. A similar CO pattern is also seen in the 3D GCM of Navarro et al. (2021) and Gilli et al. (2021). In their simulations, the CO maximum shifts westward toward the morning at 85-100 km, caused by a westward flow imposed on the SS-AS circulation. Our results overall agree with their results.

SOIR has observed a statistical difference of the CO mixing ratio profiles between two terminators via solar occultation (Vandaele et al. 2016). In Figure 6 of Vandaele et al. (2016), the CO mixing ratio below 95 km is larger at the morning terminator than the evening terminator, while CO above 105 km shows a reversed pattern. Our simulation reproduces such a pattern, but the reversal of the terminator difference occurs at a lower altitude (~90 km, see Figure 4a) than in the SOIR observations (~90-110 km).

This reversed CO terminator difference originates from transition of atmospheric flows. At 80-90 km in our model, thermal tides transport CO-rich air downward on the nightside. The RSZ flows shift the CO-rich air toward the morning terminator. CO-poor air is pumped up by the upwelling branch of the tides on the dayside and is shifted toward the evening terminator. This process results in a larger CO mixing ratio in the morning than in the evening. Above 90 km, the SS-AS circulation transports CO produced on the dayside toward both terminators. Theoretically, if the dynamical pattern is symmetric about the noon, there should be no difference between the two terminators. However, the circulation from our 3D GCM simulations is asymmetric at these altitudes. For example, at 90-95 km, zonal flows at the two terminators have different amplitudes with opposite directions (Fig. 4b). This asymmetry could cause the terminator difference of the CO mixing ratio above 90 km. The wind pattern from the GCM in Gilli et al. (2021) is also asymmetric above 110 km due to perturbations of gravity waves. The CO observations by SOIR do not show a large difference between terminators until above 120 km (Vandaele et al. 2016). This may imply that only above 120 km, the asymmetric wind pattern becomes significant enough to affect CO local-time patterns.

The reversal altitude of the CO terminator difference might be closely related to the transition from RSZ flow to SS-AS circulation on Venus. That our simulated reversal level is lower than in the SOIR observations might imply that the transition from RSZ flow to SS-AS circulation occurs at a lower level in our GCM simulations than that in the real Venusian atmosphere. Because the transition level could also vary with time and space, future observations of CO distributions are useful to constrain the flow pattern transition in the upper atmosphere of Venus.

### 3.3 $H_2O$, HCl, ClO, OCS and SO

$H_2O$ distributes almost uniformly over local time and altitude in our simulations (Figure 5a). This is because $H_2O$ is a long-lived species (Figure 5b-c). Due to thermal tides, $H_2O$ exhibits small local-time variations in the upper cloud region (58-70 km), and the amplitudes of these variations are generally less than 30 percent. The uniform distribution of $H_2O$ over local time is consistent with observations by both SPICAV (e.g., Fedorova et al. 2008) and TEXES (e.g., Encrenaz et al. 2020). SOIR observations also show no significant difference of $H_2O$ between morning and evening terminators in the upper mesosphere (Chamberlain et al. 2020).

HCl, like $H_2O$, has a long chemical lifetime and distributes uniformly over space (Figure 6). Its vertical profile in our simulations, like in previous 1D models (e.g., Yung & Demore 1982), shows a weak decrease from the cloud top to above 90 km. This sismulated profile disagrees with JCMT (James Clerk Maxwell Telescope) observations (Sandor and Clancy 2012, 2017), which show a large decrease as altitude increases. Our model seems to support the conclusion of Sandor and Clancy (2017) that the large decrease of HCl mixing ratio observed by JCMT does not originate from the SS-AS circulation. However, note that SOIR observed that HCl mixing ratio increases as altitude increases (Mahieux et al. 2015), which disagrees with the JCMT observations and also our model (and previous models). The SOIR observation suggests a chlorine source at high altitude, but no chemical hypothesis could support this source. Future observations are needed to further investigate the discrepancy among models and observations.

ClO is a short-lived species except at 80-95 km on the nightside (Figure 7). In the entire mesosphere, ClO mixing ratio is rather small, mostly < 1 ppb. But at 80-95 km on the nightside, where ClO chemical lifetime is longer, ClO can reach a few tens of ppb (Figure S6). Our simulated nightside ClO is roughly consistent with the results in a 1D nightside model from Krasnopolky (2013) but the abundance is much larger than the 1D diurnal-mean photochemical model results from Zhang et al. (2012) and Krasnopolsky (2012). Sandor and Clancy (2018) observed the nighttime ClO using JCMT and retrieved a few ppb of ClO above 85 km, which is an order of mangitude smaller than our simulated ClO mixing ratio on the nightside. Because the observed HCl from JCMT is also smaller than our simulated HCl in the upper atmosphere, we hypothesize there might be some unidentified sinks for ClO and HCl.

The vertical profile of OCS mixing ratio shows a small peak at 80-90 km. This peak is due to the downward $S_8$ flux from the top boundary in order to explain the $SO_2$ inversion. Part of the $S_8$ also converts to OCS to form a peak at 80-90 km. OCS is a short-lived chemical species on the dayside above the clouds and long-lived species on the nightside (Figure 8). Its distribution thus is largely affected by photochemistry on the dayside in the upper atmosphere and by dynamics on the nightside. OCS around 95 km exhibits a smaller mixing ratio on the dayside than the nightside and a reversed local-time pattern around 85 km (Figure S7), as a result of competition between photochemistry and dynamics. However, since OCS mixing ratios at these altitudes do not exceed 1 ppb, these local time variations are not easily observed. At ~65 km in the upper cloud, OCS mixing ratio can exceed 1 ppb, and the local time difference of the OCS mixing ratio can reach ~10 ppb. This may be an observable pattern in the future. OCS also exhibits a two-maxima local time pattern at ~65 km, similar to $SO_2$. But the larger maximum of OCS locates around the morning terminator while that of SO2 is around the evening terminator. Krasnopolsky (2010) observed a few ppb of OCS near 65 km using the CSHELL spectrograph at NASA IRTF, and indicated a pattern in which the morning OCS is more abundant than the afternoon OCS, supporting our simulated OCS local-time pattern here (Figure 10c). The OCS decrease from morning to afternoon should be related to that around 65 km, the OCS behavior is both driven by photochemistry and dynamics (Figure 8c), unlike $SO_2$, which is more driven by dynamics.

SO exhibits a complex spatial pattern (Figure 9). Since SO is a short-lived species and mainly produced by $SO_2$ photolysis, SO is more abundant on the dayside than the nightside. But in the upper cloud region on the nightside, SO has a longer chemical lifetime than the transport timescale by the RSZ flow (Figure 9c), leading to a smaller day-night contrast than that at 70-95 km. The day-night difference of SO in the upper mesosphere is consistent with the JCMT observation by Sandor et al. (2010) (Figure 10b). However, the SO mixing ratio around 95 km is lower in our model than the SPICAV observations (Belyaev et al. 2012). The SO mixing ratio shows a very strong local-time dependence (Figure 9a). Therefore, only observing the terminator SO is insufficient to understand the SO behavior. To better understand the sulfur cycle in the upper mesosphere of Venus, observations covering multiple local times on both dayside and nightside are required.

## 4. SENSITIVITY TEST

We conduct sensitivity tests to explore the effects of the horizontal diffusion coefficient $K_{xx}$, the vertical diffusion coefficient $K_{zz}$, and the horizontal resolution on our results. For simplicity, we still assume the diffusion coefficients $K_{xz}$ and $K_{zx}$ as zero.

Our sensitivity tests show that $K_{zz}$ augment does not affect the overall local-time patterns of all species discussed above (Figure S2-8). For example, the two-maximum pattern of $SO_2$ at ~64 km is still well produced in the cases with a larger $K_{zz}$ (Figure S2a). The major effect of increasing $K_{zz}$ is to increase the $SO_2$ mixing ratio below 80 km. The increase below 80 km is due to more diffusion from the lower sulfur reservoir (at ~58 km in our model). As a result, the mixing ratio of SO—a photochemical product of $SO_2$—below 80 km also increases. OCS is also sensitive to $K_{zz}$ value. As $K_{zz}$ increases, the amplitude of the OCS local-time variation at ~65 km increases despite the qualitative pattern unchanged (Figure

S7a). This implies that the OCS local-time pattern at ~65 km can be a good indicator of the strength of atmospheric vertical mixing.

Changing $K_{xx}$ from $10^9\ cm^2 s^{-1}$ (the value in the nominal case) by a factor of 10 does not affect the local-time patterns of all species discussed above (Figure S2-S8). It exerts almost no effect on the mixing ratios of the species. This is because horizontal transport by eddies only contributes a small proportion to chemical transport compared to the meridionally-mean zonal wind; the horizontal diffusion timescale is ~10$^7$-10$^9$ s (estimated by $L^2/K_{xx}$, where $L$ is planetary raidus) compared to the advection timescale of ~10$^4$ s in the upper cloud (Figure 1e). Our test also shows that increasing the horizontal resolution from 12 degrees to 6 degrees does not change the local-time patterns of the species discussed in this work (Figure S2-S8).

## 5. CONCLUSION AND DISCUSSIONS

In this paper, we investigated the local-time dependence of chemical species in the Venusian mesosphere. We used a 3D GCM and a 2D CTM to simulate species' local-time distributions and investigate the underlying mechanisms. Our models reproduce the observed local-time patterns of many chemical species such as $SO_2$ and CO. Dynamics and photochemistry play different roles in controlling the local-time patterns for different chemical species in the Venusian atmosphere.

As observed by TEXES, the local-time pattern of the $SO_2$ at ~64 km features two local maxima around terminators (Encrenaz et al. 2020). Using our model, we found that this feature is caused by the superposition of the semidiurnal thermal tide and the RSZ flow in the upper cloud. The two upwelling branches of the semidiurnal tide produce two local $SO_2$ maxima, and the superrotating wind advects the maxima toward terminators. $SO_2$ above 85 km has a large day-night difference with more $SO_2$ on the nightside, due to both chemistry and dynamics; $SO_2$ on the dayside is destroyed by photolysis, while $SO_2$ on the nightside is enriched by downwelling motions. This day-night difference of $SO_2$ in our model agrees with SPICAV occultation observations.

Circulation patterns control the CO local-time pattern over photochemical processes in the upper mesosphere. Above 80 km, CO increases as altitude increases. The upwelling of SS-AS circulation transports the CO-poor air on the dayside, while the downwelling does the opposite on the nightside. This circulation pattern decreases CO on the dayside and increases CO on the nightside. The CO local-time maximum shifts westward from midnight to the morning as altitude decreases in the upper mesosphere. This shift is consistent with microwave observations and is due to the transition from the SS-AS circulation to the RSZ flow. Below 80 km, the CO mixing ratio is nearly constant over space due to its long chemical loss timescale.

Our models also explains the CO terminator difference observed by SOIR. CO at the morning terminator is more abundant than that at the evening terminator at lower altitudes, while this pattern is reversed at higher altitudes. The difference at lower altitudes is due to thermal tides combined with the RSZ flows. The difference at higher altitudes might relate to the zonally asymmetric circulation. The reversal level simulated by our models is lower than the SOIR observations. This could indicate that the transition level from RSZ flow to SS-AS circulation in our GCM is lower than that in the Venusian atmosphere. The CO local-time variability could thus be used to constrain the atmospheric circulation of Venus.

$H_2O$ and HCl are long-lived like CO and distribute almost uniformly over both local time and altitude. The uniform distribution of $H_2O$ is qualitatively consistent with the TEXES observations. HCl vertical profiles simulated by our models disagree with JCMT observations and support that SS-AS circulation is unlikely to produce the large decrease of HCl in the upper mesosphere. ClO shows a maximum at 80-95 km on the nightside. OCS is observable in the upper cloud and also exhibits a two-maxima local-time pattern in the upper cloud. SO is a short-lived species whose mixing ratio is larger on the dayside than the nightside.

The disagreement of RSZ-to-SS-AS transition level between the model and the SOIR data needs further investigation. This transition occurs where the semidiurnal tides dissipate in the upper mesosphere. The thermal tidal waves transport retrograde angular momentum downwards to the superrotation region and decelerate the atmosphere above (Mendonca & Read 2016). These waves dissipate/break in the upper layers by radiative damping. Improving the representation of gas absorbers in the upper atmospheric region of the 3D simulations and moving the top of the model domain to higher altitudes might help reduce the disagreement in the RSZ-to-SS-AS transition altitude between the data and the model. The latter will mitigate the inaccuracies due to the top rigid model boundary, which may impact the atmospheric flow in the transition region. Also, moving the top boundary to a higher altitude will diminish the impact of the sponge layer scheme in the model's uppermost layers in the GCM.

Furthermore, in the future, new observations from the Venus missions (DAVINC+, VERITAS and EnVision) will reveal more spatial and temporal variabilities of chemical species on Venus. To understand these variabilities, the 3D GCM + 3D CTM approach could be a better way than our current approach despite a more expensive computational cost. A future 3D GCM + 3D CTM model set will show how 3D circulations (including meridional circulations) and photochemistry together control species' variabilities in the middle atmosphere of Venus.

**Acknowledgments**

This work is supported by NSF grant AST1740921 to Xi Zhang. Wencheng D. Shao is supported by the China Scholarship Council Fellowship. We thank Carver J. Bierson for the discussion of the 2D CTM. We also acknowledge use of the lux supercomputer at UC Santa Cruz, funded by NSF MRI grant AST 1828315. We thank two anonymous reviewers for constructive comments on the manuscript.

**Appendix.A Meridionally-mean chemical transport equation**

Our CTM uses the log-pressure-longitude coordinate to solve the continuity equation. From Chapter 9 and 10 in Andrews et al. (1987), the continuity equation for volume mixing ratio $\chi$ of a minor species is

$$(\rho_0 \chi)_t + \frac{(\rho_0 \chi u)_\lambda + (\rho_0 \chi v \cos\phi)_\phi}{a \cos\phi} + (\rho_0 \chi w)_z = \rho_0 S. \quad (A1)$$

Here $\rho_0 = \rho_s \exp\{-z/H\}$ is the reference background density, and $\rho_s$ and $H$ are the density at a reference level (bottom boundary) and a characteristic scale height that does not vary with height. $t$ is time. $\lambda$, $\phi$, and $z$ are longitude, latitude, and height in the log-pressure coordinate, respectively, and $u$, $v$, and $w$ are velocities in three directions. $a$ is the planetary radius. $\rho_0 S$ represents chemical production and loss rates. Subscript represents the partial derivative with respect to each coordinate.

Multiplying equation (A1) by $\cos\phi \, d\phi$, integrating it from one pole to the other pole over the meridional direction, and dividing it by $\int_{-\pi/2}^{\pi/2} \cos\phi \, d\phi$, we get

$$(\rho_0 \overline{\chi})_t + \frac{\left(\rho_0 \overline{\chi u^*}\right)_\lambda}{a} + (\rho_0 \overline{\chi w})_z = \rho_0 \overline{S}, \quad (A2)$$

where the overbar represents the average of any quantity $x$ over latitude

$$\overline{x} = \frac{\int_{-\pi/2}^{\pi/2} x \cos\phi \, d\phi}{\int_{-\pi/2}^{\pi/2} \cos\phi \, d\phi}, \quad (A3)$$

and $u^* = u/\cos\phi$.

Doing the same operation to the continuity equation for the background atmosphere

$$\frac{(\rho_0 u)_\lambda + (\rho_0 v \cos\phi)_\phi}{a \cos\phi} + (\rho_0 w)_z = 0, \quad (A4)$$

we can get a similar expression

$$\frac{\left(\rho_0 \overline{u^*}\right)_\lambda}{a} + (\rho_0 \overline{w})_z = 0. \quad (A5)$$

Combining equations (A2) and (A5) and using

$$\overline{x_1 x_2} = \overline{x_1} \, \overline{x_2} + \overline{x_1' x_2'}, \quad (A6)$$

where $x_1$ and $x_2$ are any two quantities, and $x_i' = x_i - \overline{x_i}$ ($i = 1, 2$) is the deviation

from the mean, we get

$$(\rho_0 \bar{\chi})_t + \frac{\rho_0 \overline{u^*(\bar{\chi})_\lambda}}{a} + \rho_0 \bar{w}(\bar{\chi})_z = \rho_0 \bar{S} - \left\{ \frac{\left(\rho_0 \overline{\chi' u^{*'}}\right)_\lambda}{a} + \left(\rho_0 \overline{\chi' w'}\right)_z \right\}. \quad (A7)$$

We can parameterize the deviation term in the curly bracket as diffusion:

$$(\rho_0 \bar{\chi})_t + \frac{\rho_0 \overline{u^*(\bar{\chi})_\lambda}}{a} + \rho_0 \bar{w}(\bar{\chi})_z$$
$$= \rho_0 \bar{S} - \left\{ \left( \frac{\rho_0}{a^2} K_{xx} \bar{\chi}_\lambda + \frac{\rho_0}{a} K_{xz} \bar{\chi}_z \right)_\lambda + \left( \rho_0 K_{zz} \bar{\chi}_z + \frac{\rho_0}{a} K_{zx} \bar{\chi}_\lambda \right)_z \right\}. \quad (A8)$$

This is the tracer continuity equation in the log-pressure-longitude plane, derived based on the traditionally defined longitude-latitude coordinate.

### Appendix.B Fast chemical cycles

When calculating the chemical lifetimes of $SO_2$, $SO$ and $CO$, we exclude some fast chemical cycles. We list these cycles here for a reference.

There are two fast cycles in the $SO_2$-related network. One involves species $Cl_2$:

$$\left. \begin{array}{c} 2Cl + 2SO_2 + 2M \rightarrow 2ClSO_2 + 2M \\ 2ClSO_2 \rightarrow Cl_2 + 2SO_2 \\ Cl_2 \rightarrow 2Cl \end{array} \right\}. \quad (B1)$$

The other involves species $SO$:

$$\left. \begin{array}{c} Cl + SO_2 + M \rightarrow ClSO_2 + M \\ SO + ClSO_2 \rightarrow OSCl + SO_2 \\ OSCl + M \rightarrow Cl + SO + M \end{array} \right\}. \quad (B2)$$

For $SO$, besides the cycle (B2), there is another fast cycle involving the $SO$ dimer:

$$\left. \begin{array}{c} 2SO + M \rightarrow (SO)_2 + M \\ (SO)_2 + M \rightarrow 2SO + M \end{array} \right\}. \quad (B3)$$

For $CO$, there are two fast cycles:

$$\left. \begin{array}{c} Cl + CO + CO_2 \rightarrow ClCO + CO_2 \\ ClCO + CO_2 \rightarrow Cl + CO + CO_2 \end{array} \right\}, \quad (B4)$$

and

$$\left. \begin{array}{c} Cl + CO + N_2 \rightarrow ClCO + N_2 \\ ClCO + N_2 \rightarrow Cl + CO + N_2 \end{array} \right\}. \quad (B5)$$

In Figures 1, 3 and 9, we have shown the chemical lifetime calculations for $SO_2$, CO and SO with these fast cycles excluded. Figure B1 shows the chemical lifetime calculations when these cycles are included. Comparing Figure B1 to Figures 1, 3 and 9, we found that including these cycles when calculating the chemical lifetime would give unreasonable results. For example, CO appears short-lived in the upper cloud on the dayside (panel e of Figure B1).

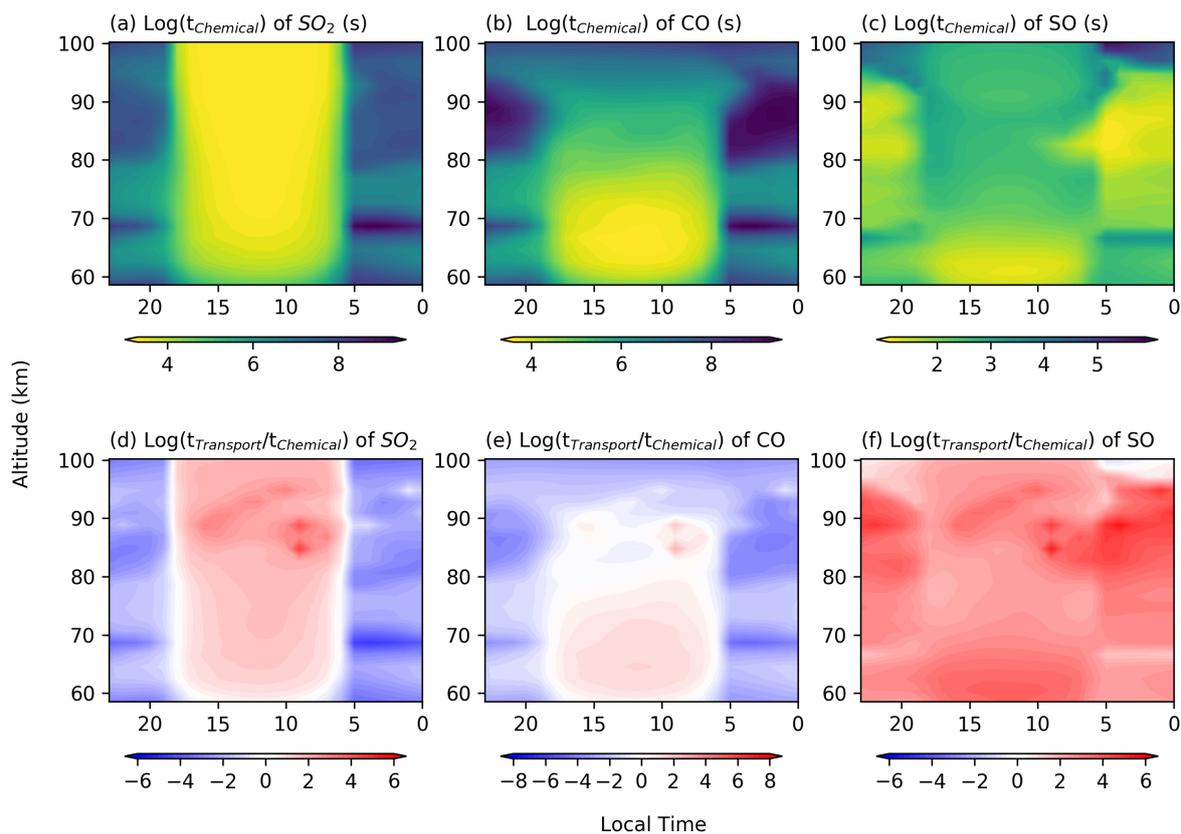

**Figure B1** Chemical lifetimes of (a) $SO_2$, (b) CO and (c) SO if the fast cycles in Appendix B are included. Panels d-f are the ratios of $t_{Transport}$ to $t_{Chemical}$ for the three species when the fast cycles are included.

**Tables**

**Table 1** Boundary conditions for several important species in the 2D CTM.

| Species | Lower Boundary Condition | Upper Boundary Condition |
|:---:|:---:|:---:|
| $SO_2$ | $f = 1.0\ ppm$ | $\phi = 0$ |
| $H_2O$ | $f = 1.0\ ppm$ | $\phi = 0$ |
| CO | $f = 45\ ppm$ | $\phi = 0$ |
| NO | $f = 5.5\ ppb$ | $\phi = 0$ |
| HCl | $f = 0.4\ ppm$ | $\phi = 0$ |
| $CO_2$ | $f = 0.965$ | $\phi = 0$ |
| OCS | $f = 1.0\ ppm$ | $\phi = 0$ |
| $S_8$ | $v = v_m$ | $\phi = -6.0 \times 10^7\ cm^{-2}\ s^{-1}$ |

Note: $f$ means the fixed volume mixing ratio, $\phi$ means the diffusive boundary flux, and $v$ is the deposition velocity. Values here are referred to those in Zhang et al. (2012) and Bierson and Zhang (2020). Species not specified here all have $\phi = 0$ at the upper boundary and the maximum deposition velocity $v_m$ (see Zhang et al. 2012) at the lower boundary (58 km).

**Table 2** Observations used in this paper.

| Observation | Altitude, km | Species | Mixing ratio range | Reference |
|---|---|---|---|---|
| TEXES | ~64 km | $SO_2$ | 150-400 ppb | Encrenaz et al. (2020) |
| JCMT | 70-100 km | | 0-90 ppb | Sandor et al. (2010) |
| SPICAV | 95-100 km | | 50-200 ppb | Belyaev et al. (2017) |
| Microwave | 80-100 km | CO | 30-1000 ppm | Clancy and Muhleman (1985) |
| JCMT | 75-100 km | | 50-1000 ppm | Clancy et al. (2003) |
| SOIR | 85-130 km | | $10^{-4}$-$10^{-1}$ | Vandaele et al. (2016) |
| TEXES | ~64 km | $H_2O$ | ~1 ppm | Encrenaz et al. (2020) |
| JCMT | 70-100 km | HCl | 0-450 ppb | Sandor and Clancy (2012,2017) |
| SOIR | 70-105 km | | 30-800 ppb | Mahieux et al. (2015) |
| JCMT | 70-100 km | ClO | 1.5-3.7 ppb | Sandor and Clancy (2018) |
| CSHELL | ~65 km | OCS | 0.3-9 ppb | Krasnopolsky (2010) |
| JCMT | 70-100 km | SO | 0-30 ppb | Sandor et al. (2010) |
| SPICAV | 85-105 km | | 10-1000 ppb | Belyaev et al. (2012) |

**Figures**

**Figure 1**

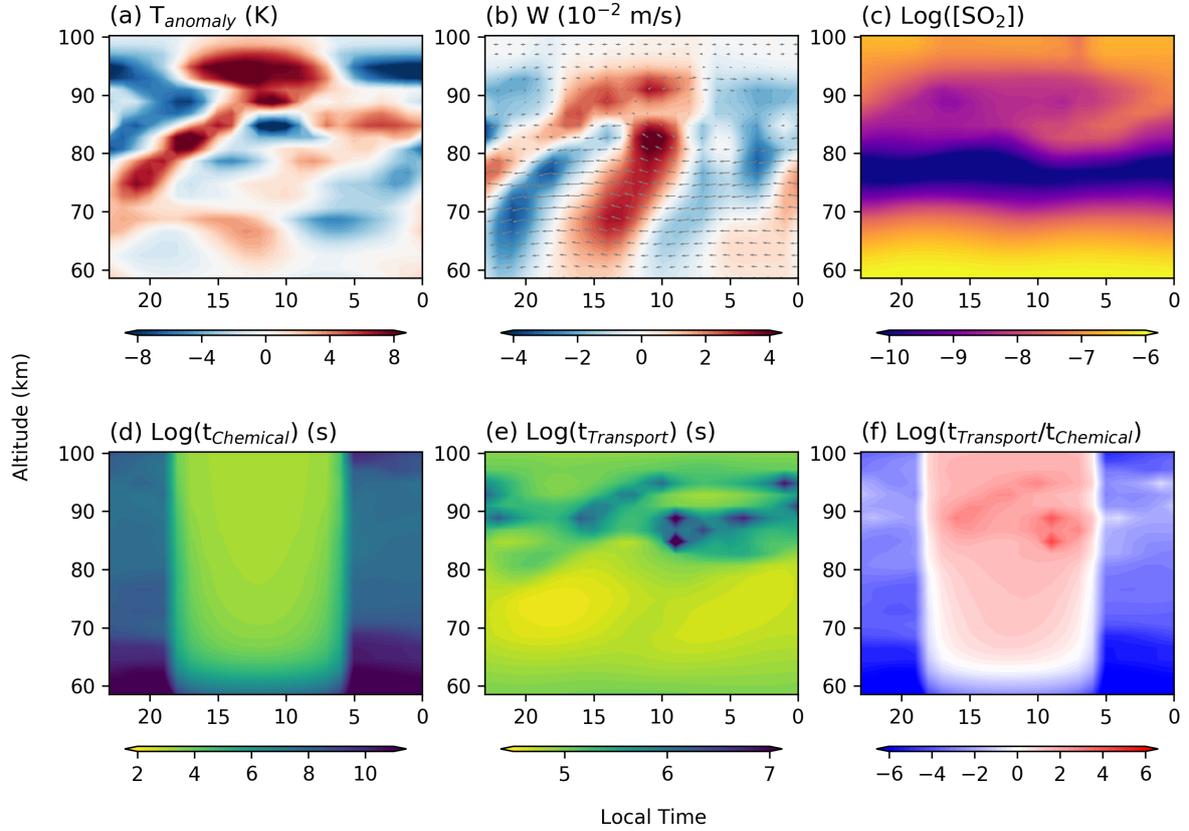

**Figure 1** Local-time dependence of (a) temperature anomalies $T_{anomaly}$, (b) vertical velocity W, (c) $SO_2$ mixing ratio, (d) $SO_2$ chemical loss timescale $t_{Chemical}$, (e) transport timescale $t_{Transport}$, and (f) ratio of $t_{Transport}$ to $t_{Chemical}$. The wind field (m/s) is superposed on panel b. Temperature and wind fields are from the OASIS simulations (Mendonça & Buchhave 2020), and the $SO_2$ mixing ratios is from the 2D CTM. Temperature anomaly is the deviation from an average temperature profile shown in Figure S1. Note that both the 2D CTM and OASIS use (log-)pressure coordinate. Height at the vertical axis in this plot represents the isobaric level and is derived from pressure by using the VIRA model (cf. Table 1 of Mendonça & Read 2016). The local time 06:00 is the morning terminator, 18:00 the evening terminator, 12:00 the noon, and 00:00 the midnight. Earlier local time means eastward shift on Venus. The ratio of $t_{Transport}$ to $t_{Chemical}$ indicates the main driven mechanism for the species distribution: the ratio smaller than unity (blue region

in panel f) implies a mainly transport-driven regime; the ratio larger than unity (red region in panel f) implies a mainly photochemistry-driven regime; the ratio around unity (white region in panel f) implies the transition between the two regimes.

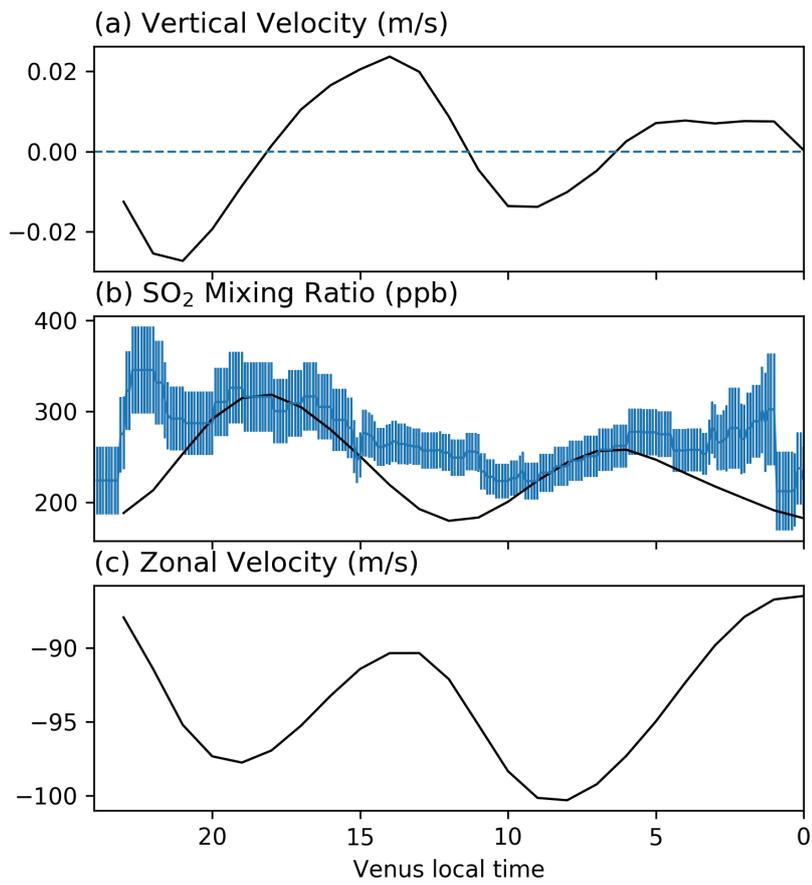

**Figure 2**

**Figure 2** Local-time distributions of (a) vertical velocity, (b) $SO_2$ mixing ratio, and (c) zonal velocity around 64 km. Observational data (error bars) in (b) are from TEXES/IRTF (Encrenaz et al. 2020). Positive vertical velocity is upward, and negative zonal velocity is westward.

**Figure 3**

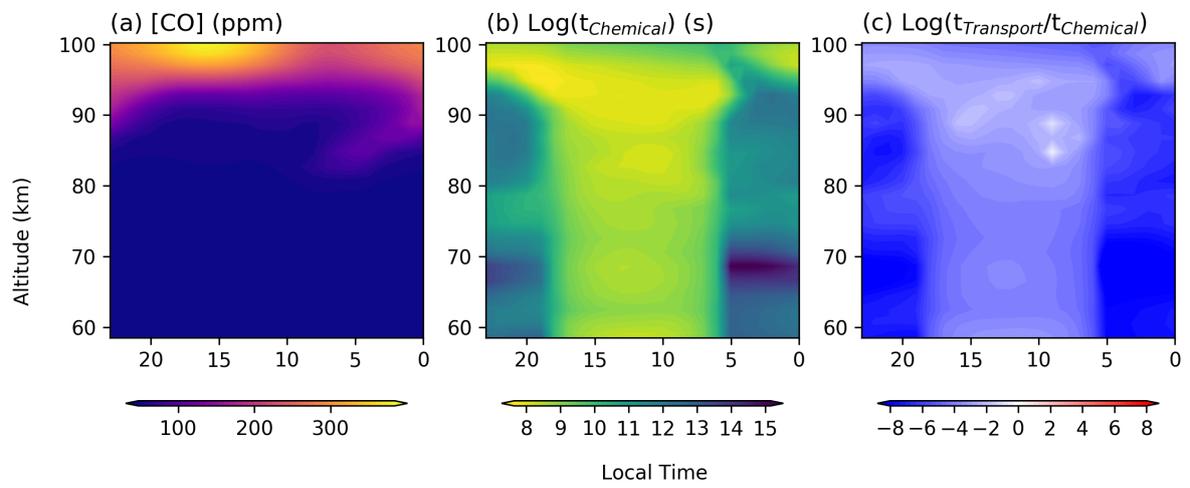

**Figure 3** Local time dependence of (a) CO mixing ratio, (b) CO chemical loss timescale $t_{Chemical}$ and (c) ratio of $t_{Transport}$ to $t_{Chemical}$.

**Figure 4**

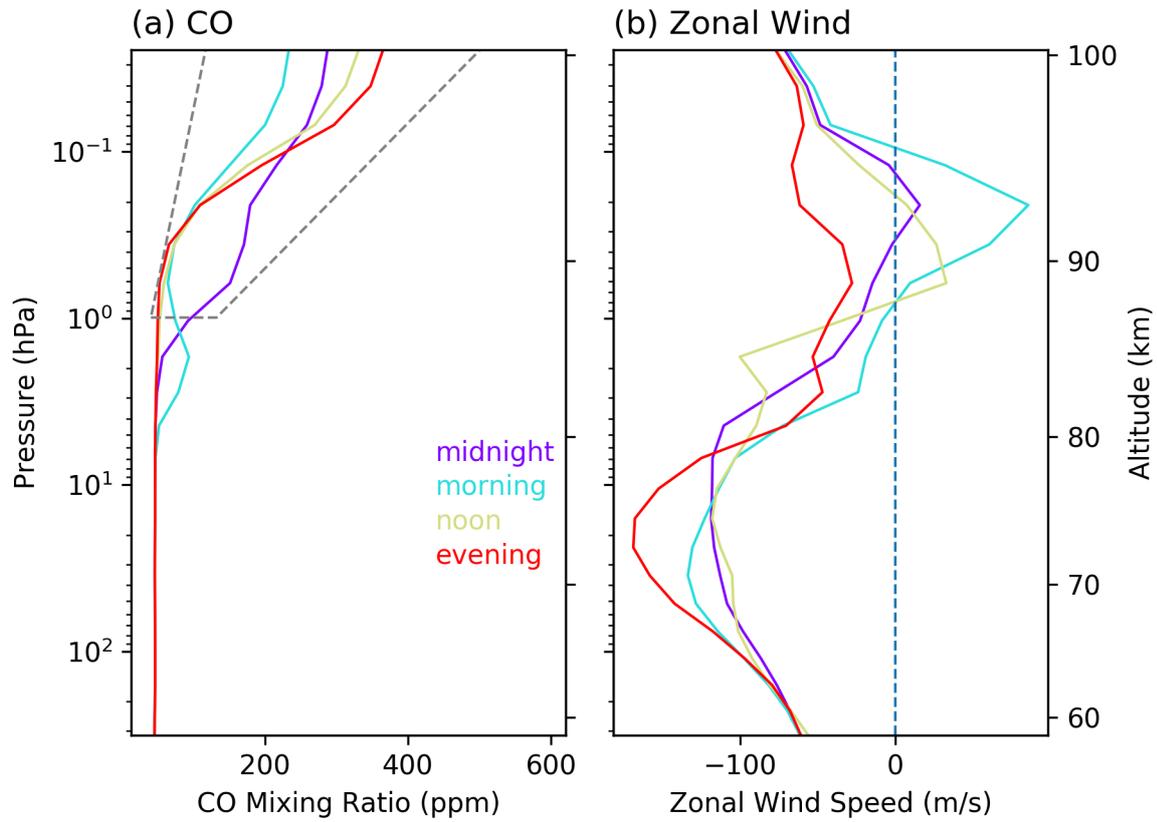

**Figure 4** Vertical profiles of (a) CO mixing ratio and (b) zonal wind at different local times. Altitude derived from the VIRA model is shown on the right axis. The grey dashed line in panel a encloses a region corresponding to a rough range of the observations by Vandaele et al. (2016). The blue dashed line in panel b is zero zonal wind.

**Figure 5**

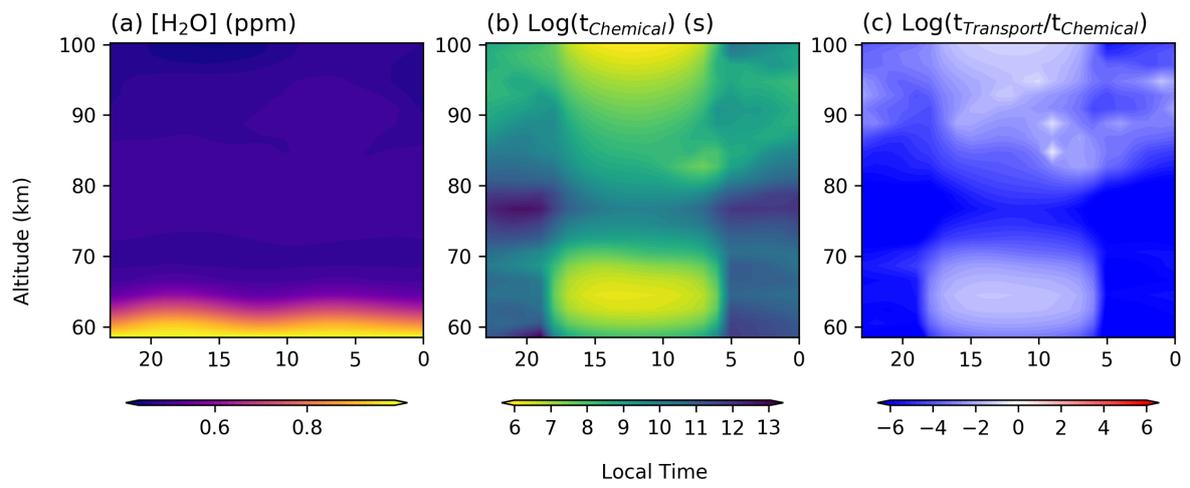

**Figure 5** Same as Figure 3, but for $H_2O$.

**Figure 6**

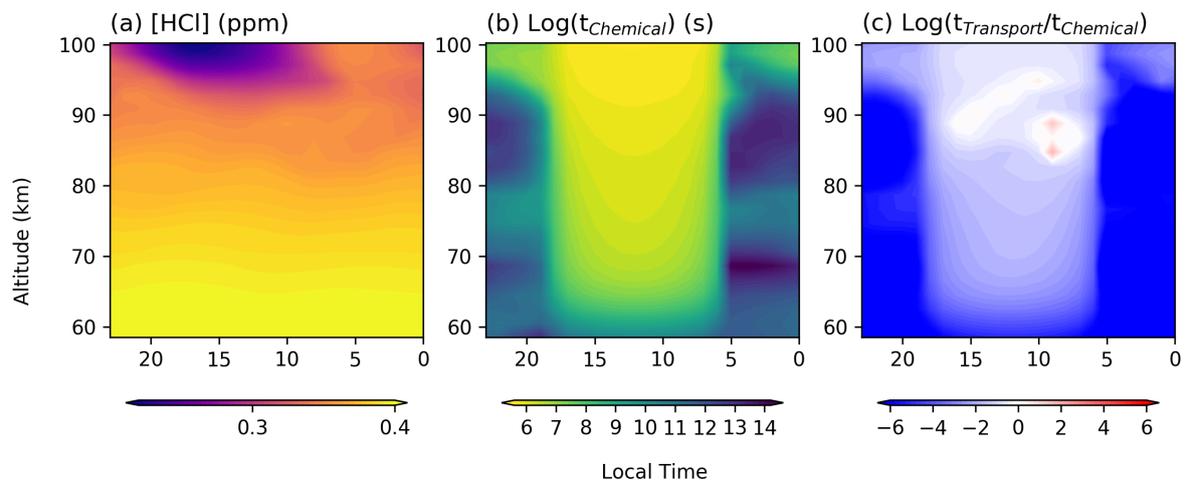

**Figure 6** Same as Figure 3, but for HCl.

**Figure 7**

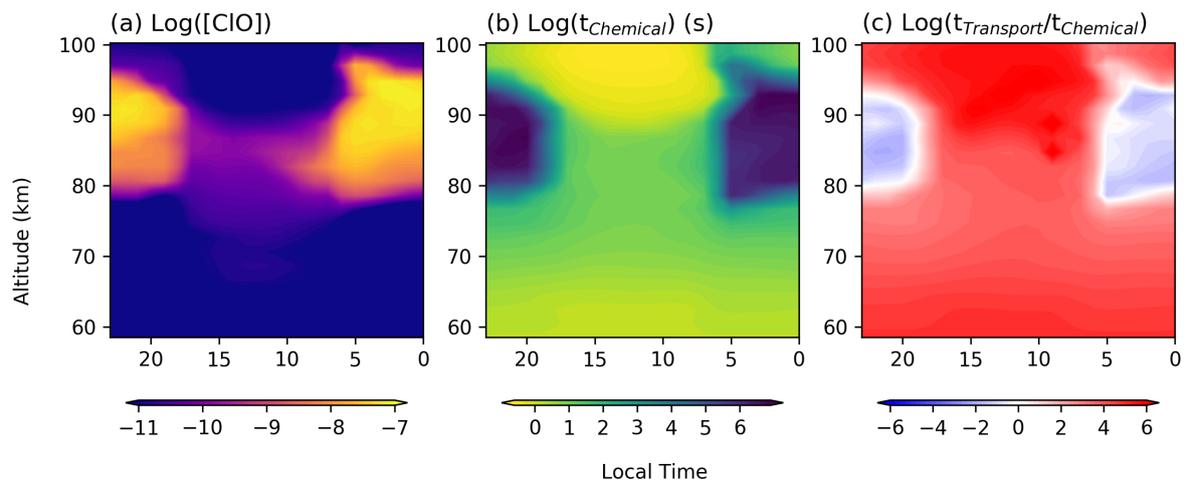

**Figure 7** Same as Figure 3, but for ClO.

**Figure 8**

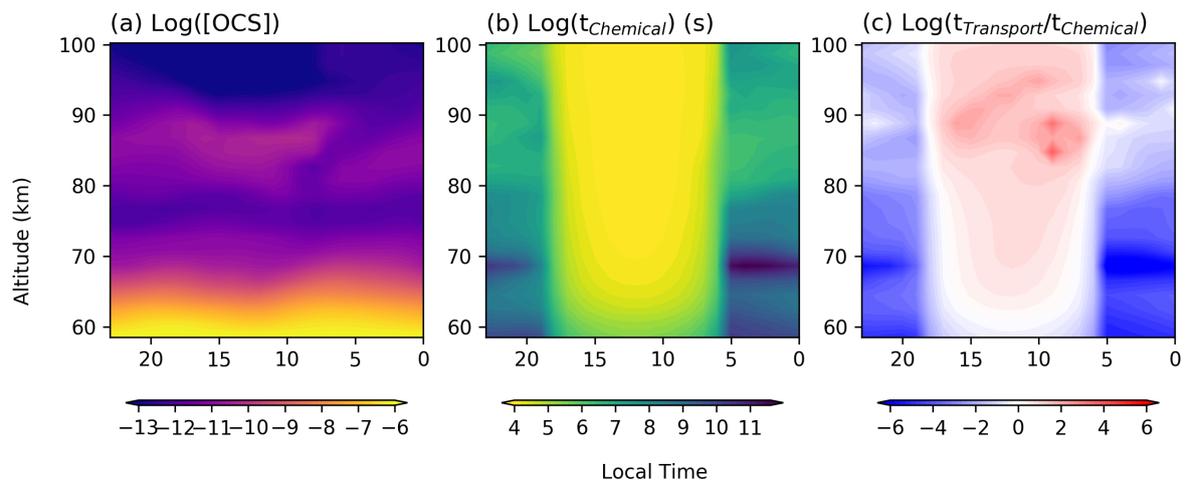

**Figure 8** Same as Figure 3, but for OCS.

**Figure 9**

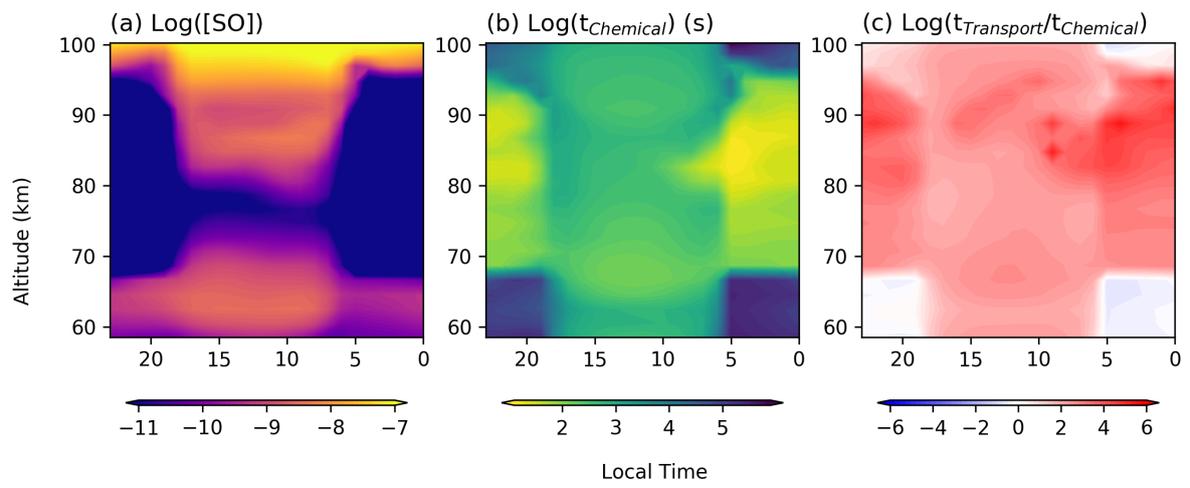

**Figure 9** Same as Figure 3, but for SO.

**Figure 10**

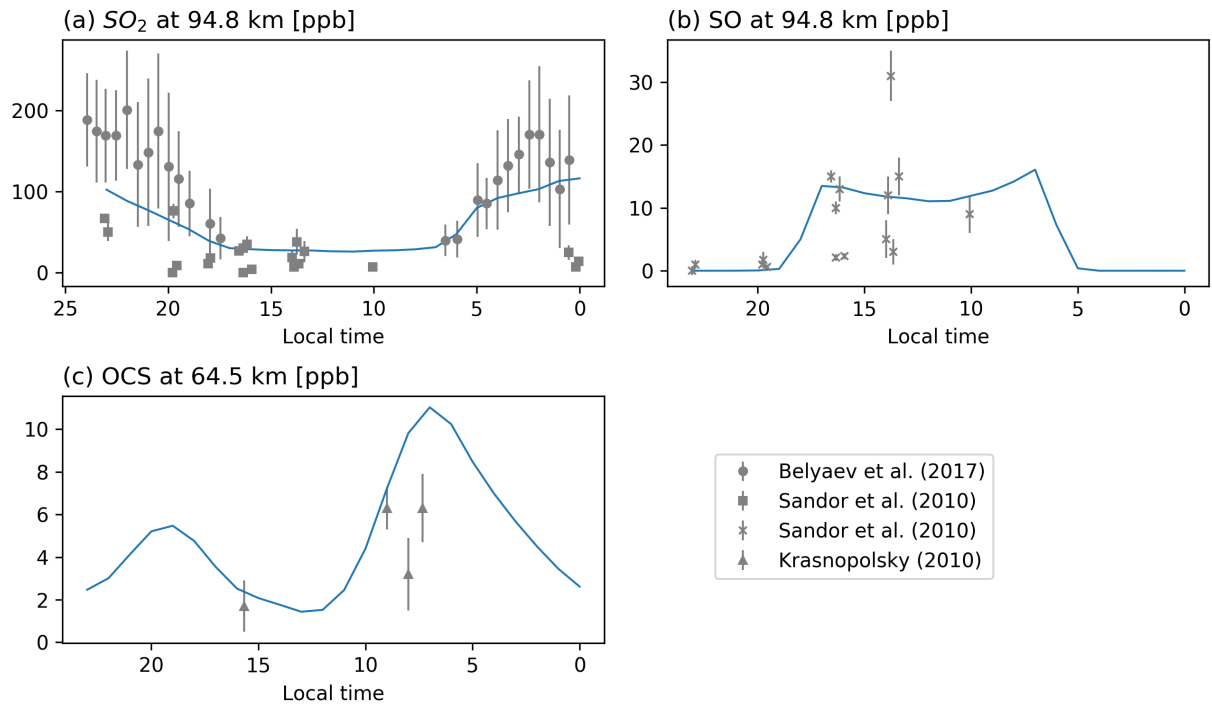

**Figure 10** Local time variations of volume mixing ratios of $SO_2$, SO and OCS from both our model and observations (figure 11b of Belyaev et al. 2017; Sandor et al. 2010; Krasnopolsky 2010). Note that the observation altitude is not necessarily exactly the value shown in this plot (refer to Table 2 to see the observation altitude range).

**Supporting Data**

**Figure S1**

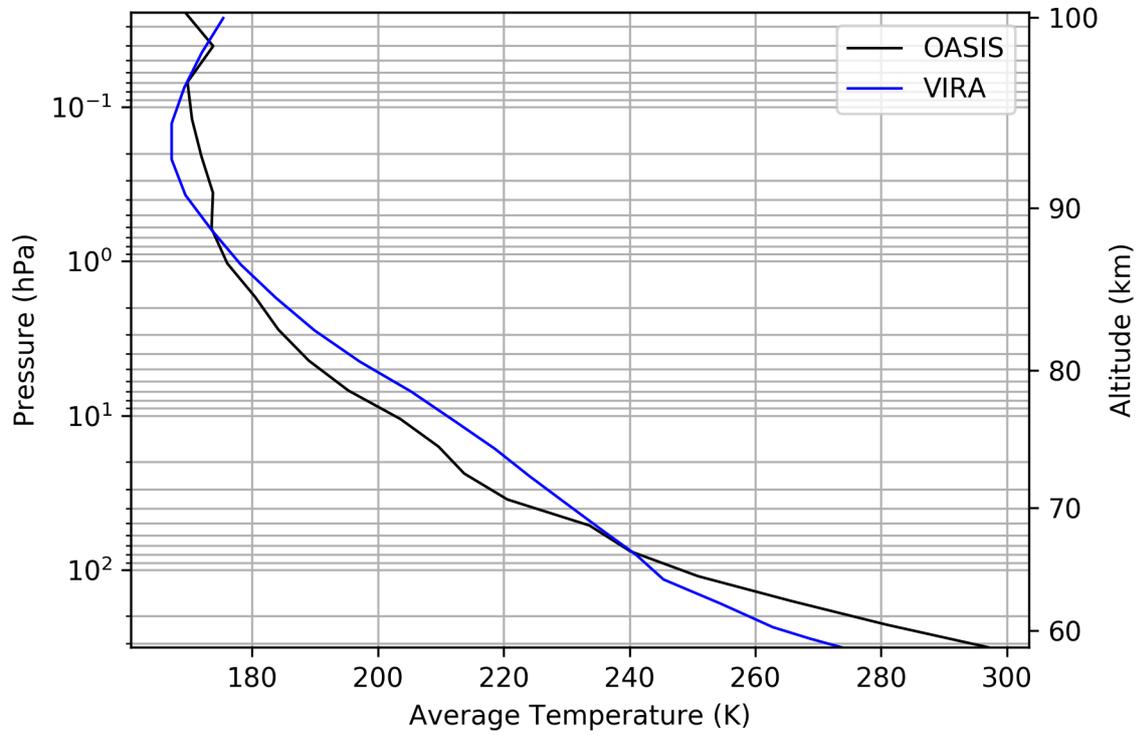

**Figure S1** Average temperature profile simulated by our GCM (black) and VIRA temperature profile (blue). The temperature anomaly in Figure 1 is the deviation from this GCM's average temperature profile.

**Figure S2**

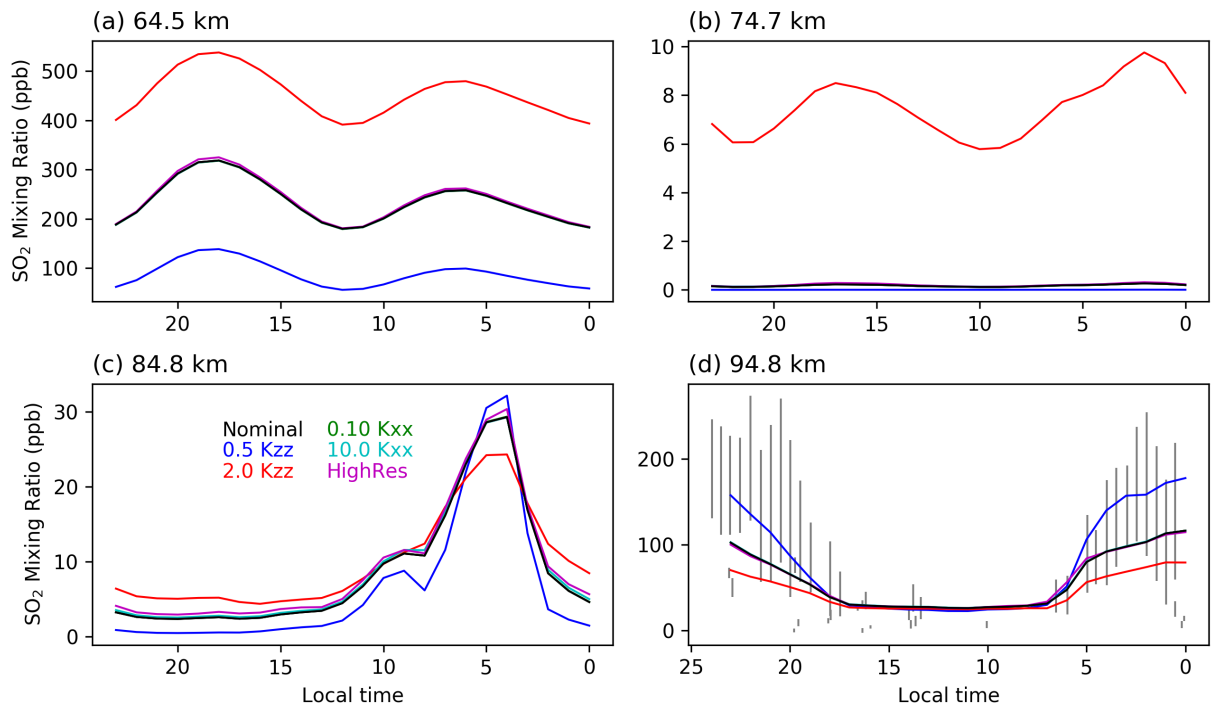

**Figure S2** Local-time distributions of $SO_2$ mixing ratio at different altitudes (a-d) for different cases: our nominal case (black); cases with Kzz enlarged by 1.5 (blue) and 2.0 (red); cases with $K_{xx}$ changed by a factor of 0.1 (green) and 10.0 (cyan); case with a higher (double) horizontal resolution (magenta). Note that green, cyan, and black lines are almost overlapping with each other. In panel d, error bars show observations at 95-100 km from SPICAV/VEx by Belyaev et al. (2017) and observations at 70-100 km from JCMT by Sandor et al. (2010).

**Figure S3**

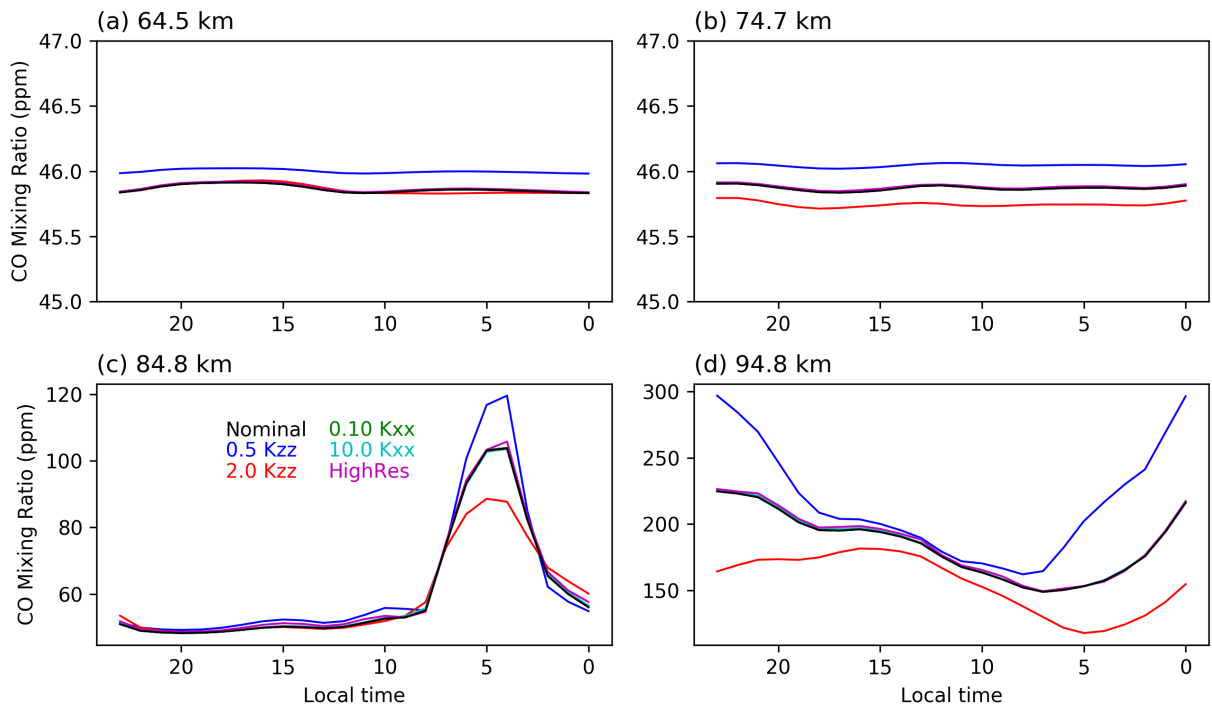

**Figure S3** Same as Figure S2 but for CO.

**Figure S4**

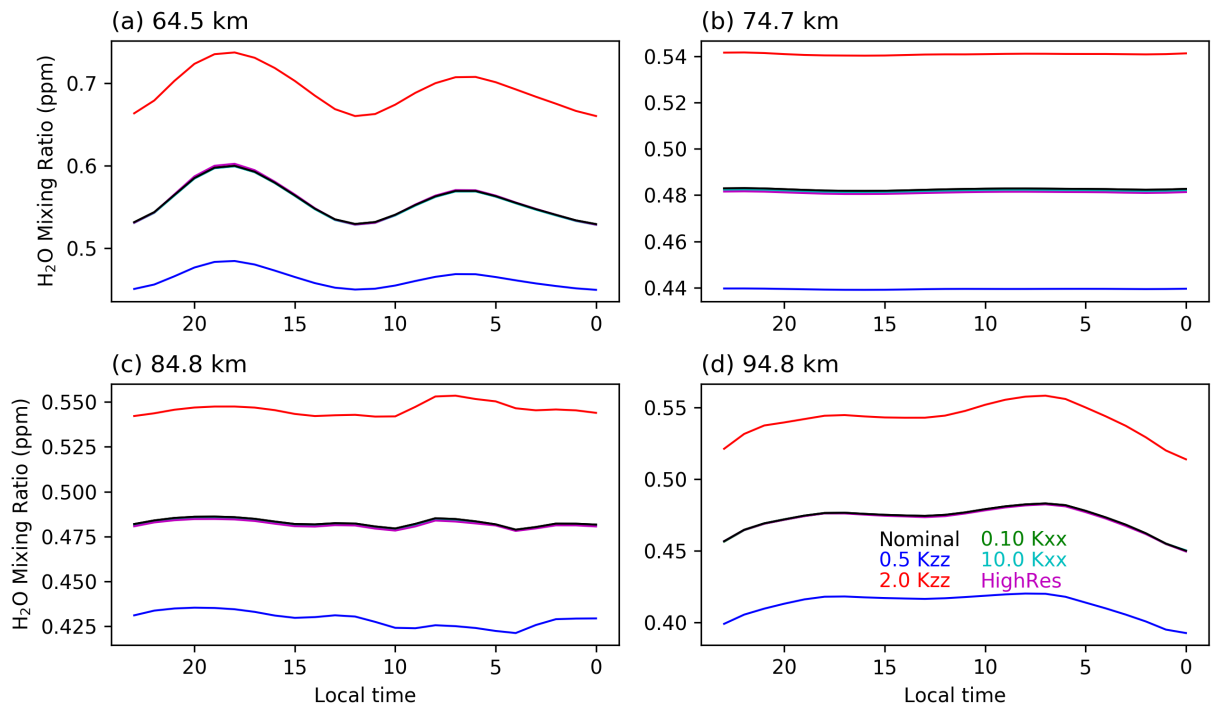

**Figure S4** Same as Figure S2 but for $H_2O$.

**Figure S5**

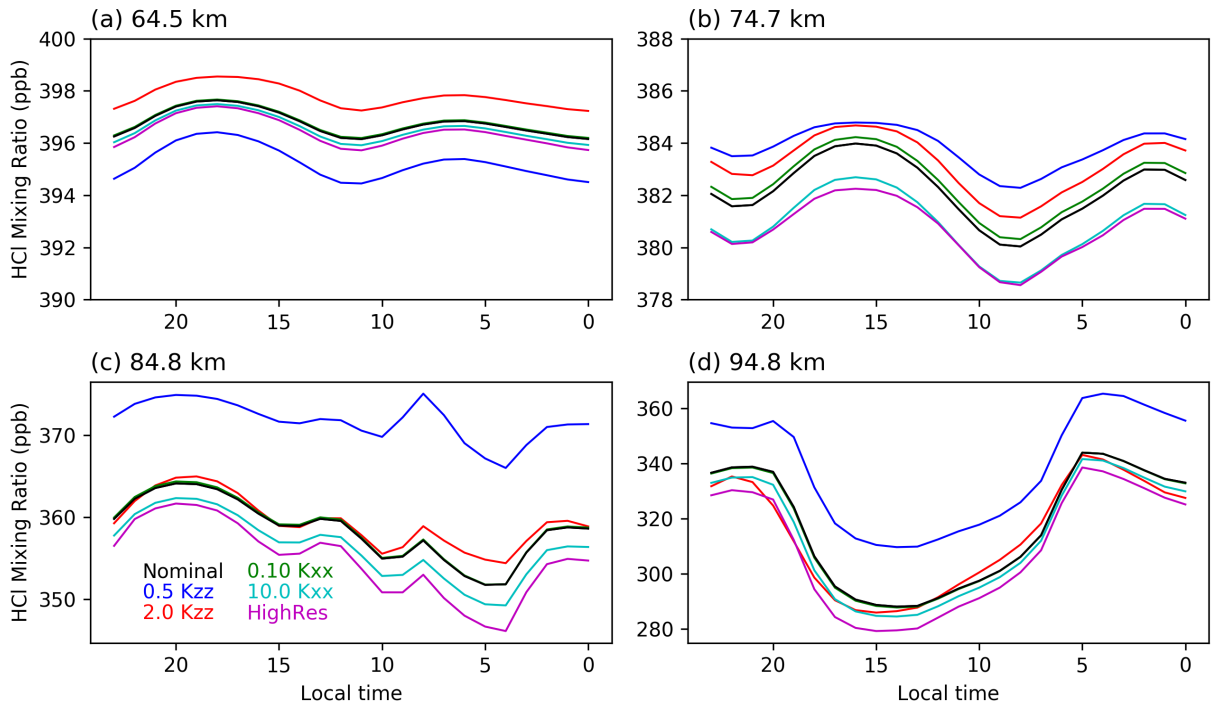

**Figure S5** Same as Figure S2 but for HCl.

**Figure S6**

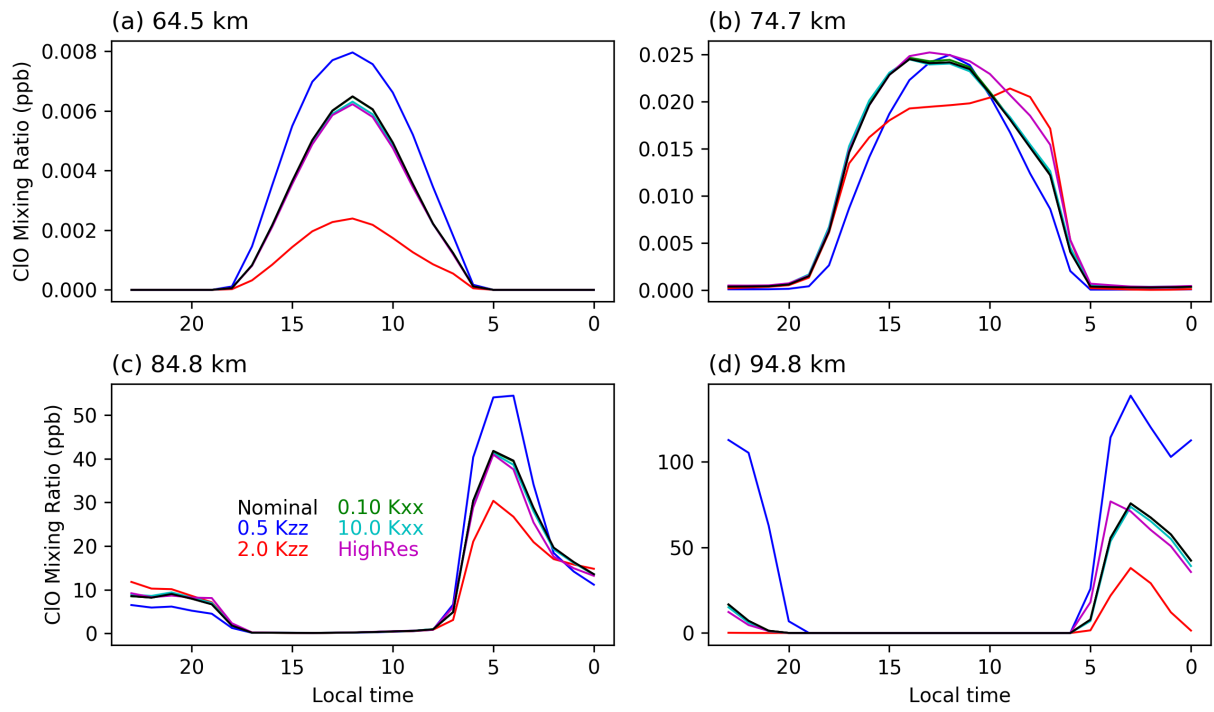

**Figure S6** Same as Figure S2 but for ClO.

**Figure S7**

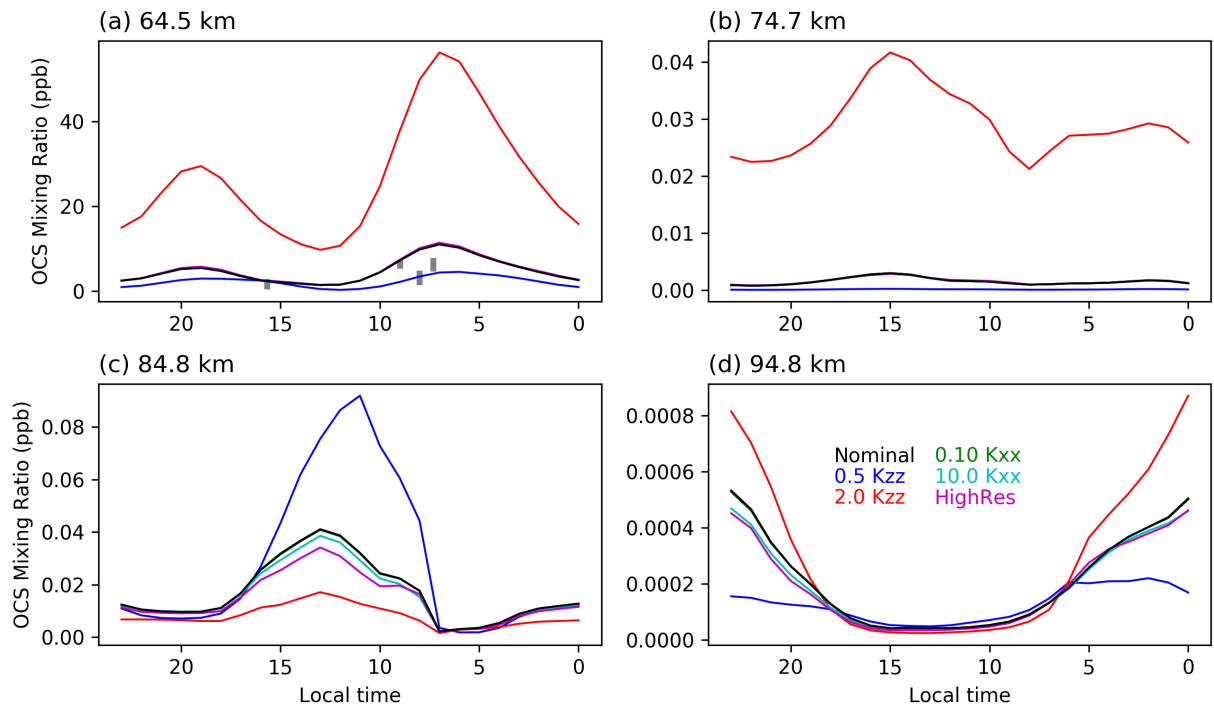

**Figure S7** Same as Figure S2 but for OCS. In panel a, grey bars show a few observation points near 65 km from CSHELL/IRTF by Krasnopolsky (2010).

**Figure S8**

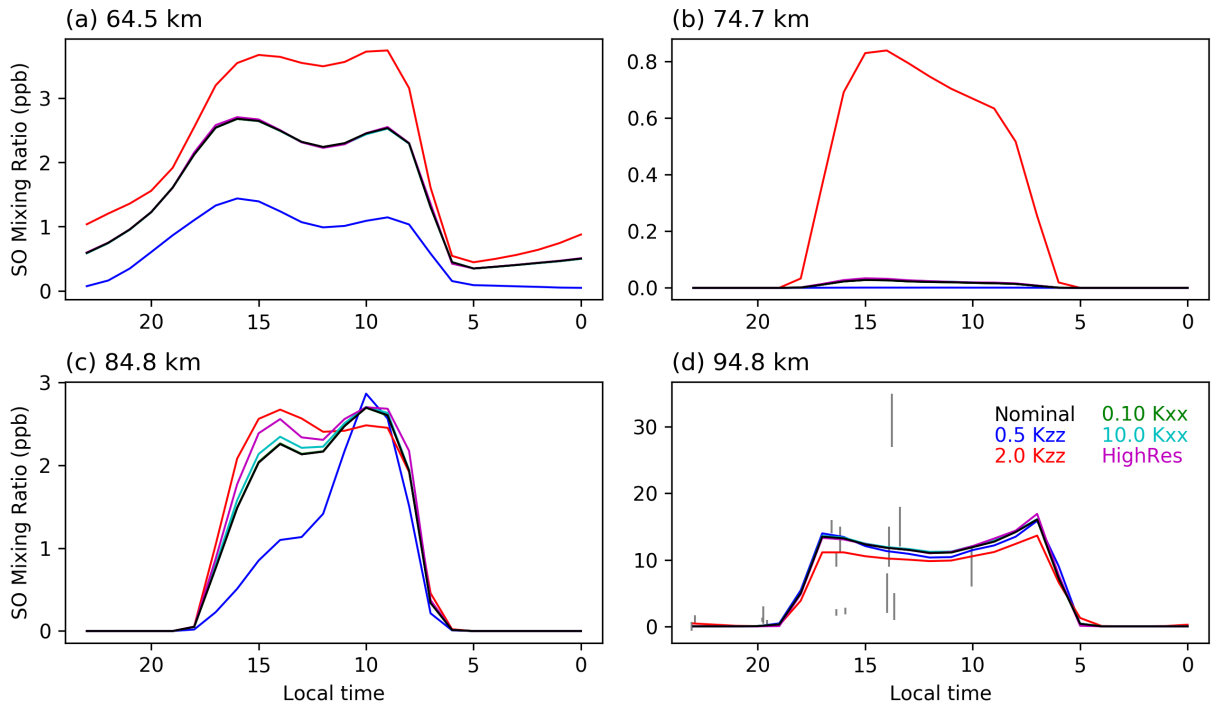

**Figure S8** Same as Figure S2 but for SO. In panel d, error bars show observations at 70-100 km from JCMT by Sandor et al. (2010).

## References

Andrews, D. G., Holton, J. R., & Leovy, C. B. (1987). *Middle Atmosphere Dynamics*. Academic Press.

Belyaev, D. A., Montmessin, F., Bertaux, J.-L., Mahieux, A., Fedorova, A. A., Korablev, O. I., Marcq, E., Yung, Y. L., & Zhang, X. (2012). Vertical profiling of SO2 and SO above Venus' clouds by SPICAV/SOIR solar occultations. *Icarus*, *217*(2), 740–751. https://doi.org/10.1016/j.icarus.2011.09.025

Belyaev, D. A., Evdokimova, D. G., Montmessin, F., Bertaux, J.-L., Korablev, O. I., Fedorova, A. A., Marcq, E., Soret, L., & Luginin, M. S. (2017). Night side distribution of SO2 content in Venus' upper mesosphere. *Icarus*, *294*, 58–71. https://doi.org/10.1016/j.icarus.2017.05.002

Bierson, C. J., & Zhang, X. (2020). Chemical cycling in the Venusian atmosphere: A full photochemical model from the surface to 110 km. *Journal of Geophysical Research: Planets*, *125*(7), e2019JE006159.

Bougher, S. W., Rafkin, S., & Drossart, P. (2006). Dynamics of the Venus upper atmosphere: Outstanding problems and new constraints expected from Venus Express. *Planetary and Space Science*, *54*(13), 1371–1380. https://doi.org/10.1016/j.pss.2006.04.023

Chamberlain, S., Mahieux, A., Robert, S., Piccialli, A., Trompet, L., Vandaele, A. C., & Wilquet, V. (2020). SOIR/VEx observations of water vapor at the terminator in the Venus mesosphere. *Icarus*, *346*, 113819. https://doi.org/10.1016/j.icarus.2020.113819

Clancy, R. T., & Muhleman, D. O. (1985). Diurnal CO variations in the Venus mesophere from CO microwave spectra. *Icarus*, *64*(2), 157–182. https://doi.org/10.1016/0019-1035(85)90084-3

Clancy, R. T., Sandor, B. J., & Moriarty-Schieven, G. H. (2003). Observational definition of the Venus mesopause: Vertical structure, diurnal variation, and temporal instability. *Icarus*, *161*(1), 1–16.

https://doi.org/10.1016/S0019-1035(02)00022-2

Crisp, D. (1986). Radiative forcing of the Venus mesosphere: I. Solar fluxes and heating rates. *Icarus*, *67*(3), 484–514. https://doi.org/10.1016/0019-1035(86)90126-0

Encrenaz, T., Greathouse, T. K., Marcq, E., Sagawa, H., Widemann, T., Bézard, B., Fouchet, T., Lefèvre, F., Lebonnois, S., & Atreya, S. K. (2019). HDO and SO2 thermal mapping on Venus-IV. Statistical analysis of the SO2 plumes. *Astronomy & Astrophysics*, *623*, A70.

Encrenaz, T., Greathouse, T. K., Marcq, E., Sagawa, H., Widemann, T., Bézard, B., Fouchet, T., Lefèvre, F., Lebonnois, S., & Atreya, S. K. (2020). HDO and SO 2 thermal mapping on Venus. V. Evidence for along-term anti-correlation. *Astronomy and Astrophysics-A&A*, in-press.

Encrenaz, T., Greathouse, T. K., Richter, M. J., DeWitt, C., Widemann, T., Bézard, B., Fouchet, T., Atreya, S. K., & Sagawa, H. (2016). HDO and SO2 thermal mapping on Venus—III. Short-term and long-term variations between 2012 and 2016. *Astronomy & Astrophysics*, *595*, A74. https://doi.org/10.1051/0004-6361/201628999

Encrenaz, T., Greathouse, T. K., Richter, M. J., Lacy, J., Widemann, T., Bézard, B., Fouchet, T., deWitt, C., & Atreya, S. K. (2013). HDO and SO2 thermal mapping on Venus—II. The SO2 spatial distribution above and within the clouds. *Astronomy & Astrophysics*, *559*, A65. https://doi.org/10.1051/0004-6361/201322264

Encrenaz, T., Greathouse, T. K., Roe, H., Richter, M., Lacy, J., Bézard, B., Fouchet, T., & Widemann, T. (2012). HDO and SO2 thermal mapping on Venus: Evidence for strong SO2 variability. *Astronomy & Astrophysics*, *543*, A153.

Fedorova, A., Korablev, O., Vandaele, A.-C., Bertaux, J.-L., Belyaev, D., Mahieux, A., Neefs, E., Wilquet, W. V., Drummond, R., Montmessin, F., & Villard, E. (2008). HDO and H2O vertical distributions and isotopic ratio in the Venus mesosphere by Solar Occultation at Infrared spectrometer on board

Venus Express. *Journal of Geophysical Research: Planets*, *113*(E5). https://doi.org/10.1029/2008JE003146

Fukuya, K., Imamura, T., Taguchi, M., Fukuhara, T., Kouyama, T., Horinouchi, T., Peralta, J., Futaguchi, M., Yamada, T., Sato, T. M., Yamazaki, A., Murakami, S., Satoh, T., Takagi, M., & Nakamura, M. (2021). The nightside cloud-top circulation of the atmosphere of Venus. *Nature*, *595*(7868), 511–515. https://doi.org/10.1038/s41586-021-03636-7

Gilli, G., Lebonnois, S., González-Galindo, F., López-Valverde, M. A., Stolzenbach, A., Lefèvre, F., Chaufray, J. Y., & Lott, F. (2017). Thermal structure of the upper atmosphere of Venus simulated by a ground-to-thermosphere GCM. *Icarus*, *281*, 55–72. https://doi.org/10.1016/j.icarus.2016.09.016

Gilli, G., Navarro, T., Lebonnois, S., Quirino, D., Silva, V., Stolzenbach, A., Lefèvre, F., & Schubert, G. (2021). Venus upper atmosphere revealed by a GCM: II. Model validation with temperature and density measurements. *Icarus*, 114432. https://doi.org/10.1016/j.icarus.2021.114432

Jessup, K. L., Marcq, E., Mills, F., Mahieux, A., Limaye, S., Wilson, C., Allen, M., Bertaux, J.-L., Markiewicz, W., Roman, T., Vandaele, A.-C., Wilquet, V., & Yung, Y. (2015). Coordinated Hubble Space Telescope and Venus Express Observations of Venus' upper cloud deck. *Icarus*, *258*, 309–336. https://doi.org/10.1016/j.icarus.2015.05.027

Jiang, X., Camp, C. D., Shia, R., Noone, D., Walker, C., & Yung, Y. L. (2004). Quasi-biennial oscillation and quasi-biennial oscillation–annual beat in the tropical total column ozone: A two-dimensional model simulation. *Journal of Geophysical Research: Atmospheres*, *109*(D16). https://doi.org/10.1029/2003JD004377

Kerzhanovich, V. V., & Limaye, S. S. (1985). Circulation of the atmosphere from the surface to 100 km. *Advances in Space Research*, *5*(11), 59–83. https://doi.org/10.1016/0273-1177(85)90198-X

Knollenberg, R. G., & Hunten, D. M. (1980). The microphysics of the clouds of Venus: Results of the Pioneer Venus Particle Size Spectrometer Experiment. *Journal of Geophysical Research: Space Physics*, *85*(A13), 8039–8058. https://doi.org/10.1029/JA085iA13p08039

Krasnopolsky, V. A. (2010). Spatially-resolved high-resolution spectroscopy of Venus 2. Variations of HDO, OCS, and SO2 at the cloud tops. *Icarus*, *209*(2), 314–322. https://doi.org/10.1016/j.icarus.2010.05.008

Krasnopolsky, V. A. (2012). A photochemical model for the Venus atmosphere at 47–112km. *Icarus*, *218*(1), 230–246. https://doi.org/10.1016/j.icarus.2011.11.012

Krasnopolsky, V. A. (2013). Nighttime photochemical model and night airglow on Venus. *Planetary and Space Science*, *85*, 78–88. https://doi.org/10.1016/j.pss.2013.05.022

Lebonnois, S., Hourdin, F., Eymet, V., Crespin, A., Fournier, R., & Forget, F. (2010). Superrotation of Venus' atmosphere analyzed with a full general circulation model. *Journal of Geophysical Research: Planets*, *115*(E6). https://doi.org/10.1029/2009JE003458

Lellouch, E., Goldstein, J. J., Rosenqvist, J., Bougher, S. W., & Paubert, G. (1994). Global Circulation, Thermal Structure, and Carbon Monoxide Distribution in Venus' Mesosphere in 1991. *Icarus*, *110*(2), 315–339. https://doi.org/10.1006/icar.1994.1125

Limaye, S. S. (2007). Venus atmospheric circulation: Known and unknown. *Journal of Geophysical Research: Planets*, *112*(E4). https://doi.org/10.1029/2006JE002814

Mahieux, A., Wilquet, V., Vandaele, A. C., Robert, S., Drummond, R., Chamberlain, S., Grau Ribes, A., & Bertaux, J. L. (2015). Hydrogen halides measurements in the Venus mesosphere retrieved from SOIR on board Venus express. *Planetary and Space Science*, *113–114*, 264–274. https://doi.org/10.1016/j.pss.2014.12.014

Marcq, E., Bertaux, J.-L., Montmessin, F., & Belyaev, D. (2013). Variations of sulphur dioxide at the cloud top of Venus's dynamic atmosphere. *Nature Geoscience*, *6*(1), 25–28.

Marcq, E., Jessup, K. L., Baggio, L., Encrenaz, T., Lee, Y. J., Montmessin, F., Belyaev, D., Korablev, O., & Bertaux, J.-L. (2020). Climatology of SO2 and UV absorber at Venus' cloud top from SPICAV-UV nadir dataset. *Icarus*, *335*, 113368.

Mendonça, J. M., & Buchhave, L. A. (2020). Modelling the 3D climate of Venus with oasis. *Monthly Notices of the Royal Astronomical Society*, *496*(3), 3512–3530. https://doi.org/10.1093/mnras/staa1618

Mendonça, J. M., & Read, P. L. (2016). Exploring the Venus global super-rotation using a comprehensive general circulation model. *Planetary and Space Science*, *134*, 1–18. https://doi.org/10.1016/j.pss.2016.09.001

Mills, F. P. (1998). *I. Observations and photochemical modeling of the Venus middle atmosphere. II. Thermal infrared spectroscopy of Europa and Callisto* [PhD Thesis]. California Institute of Technology.

Mills, F. P., & Allen, M. (2007). A review of selected issues concerning the chemistry in Venus' middle atmosphere. *Planetary and Space Science*, *55*(12), 1729–1740. https://doi.org/10.1016/j.pss.2007.01.012

Navarro, T., Gilli, G., Schubert, G., Lebonnois, S., Lefèvre, F., & Quirino, D. (2021). Venus' upper atmosphere revealed by a GCM: I. Structure and variability of the circulation. *Icarus*, 114400. https://doi.org/10.1016/j.icarus.2021.114400

Pechmann, J. B., & Ingersoll, A. P. (1984). Thermal Tides in the Atmosphere of Venus: Comparison of Model Results with Observations. *Journal of the Atmospheric Sciences*, *41*(22), 3290–3313. https://doi.org/10.1175/1520-0469(1984)041<3290:TTITAO>2.0.CO;2

Peralta, J., Luz, D., Berry, D. L., Tsang, C. C. C., Sánchez-Lavega, A., Hueso, R., Piccioni, G., & Drossart, P. (2012). Solar migrating atmospheric tides in the winds of the polar region of Venus. *Icarus*, *220*(2), 958–970. https://doi.org/10.1016/j.icarus.2012.06.015

Pollack, J. B., & Young, R. (1975). Calculations of the Radiative and Dynamical State of the Venus Atmosphere. *Journal of the Atmospheric Sciences*, *32*(6), 1025–1037. https://doi.org/10.1175/1520-0469(1975)032<1025:COTRAD>2.0.CO;2

Prather, M. J. (1986). Numerical advection by conservation of second-order moments. *Journal of Geophysical Research: Atmospheres*, *91*(D6), 6671–6681. https://doi.org/10.1029/JD091iD06p06671

Sánchez-Lavega, A., Hueso, R., Piccioni, G., Drossart, P., Peralta, J., Pérez-Hoyos, S., Wilson, C. F., Taylor, F. W., Baines, K. H., Luz, D., Erard, S., & Lebonnois, S. (2008). Variable winds on Venus mapped in three dimensions. *Geophysical Research Letters*, *35*(13). https://doi.org/10.1029/2008GL033817

Sandor, B. J., Todd Clancy, R., Moriarty-Schieven, G., & Mills, F. P. (2010). Sulfur chemistry in the Venus mesosphere from SO2 and SO microwave spectra. *Icarus*, *208*(1), 49–60. https://doi.org/10.1016/j.icarus.2010.02.013

Sandor, B. J., & Clancy, R. T. (2012). Observations of HCl altitude dependence and temporal variation in the 70–100km mesosphere of Venus. *Icarus*, *220*(2), 618–626. https://doi.org/10.1016/j.icarus.2012.05.016

Sandor, B. J., & Clancy, R. T. (2017). Diurnal observations of HCl altitude variation in the 70–100 km mesosphere of Venus. *Icarus*, *290*, 156–161. https://doi.org/10.1016/j.icarus.2017.02.017

Sandor, B. J., & Clancy, R. T. (2018). First measurements of ClO in the Venus atmosphere – Altitude dependence and temporal variation. *Icarus*, *313*, 15–24. https://doi.org/10.1016/j.icarus.2018.04.022

Shao, W. D., Zhang, X., Bierson, C. J., & Encrenaz, T. (2020). Revisiting the Sulfur-Water Chemical System in the Middle Atmosphere of Venus. *Journal of Geophysical Research: Planets*, *125*(8), e2019JE006195.

Shia, R.-L., Ha, Y. L., Wen, J.-S., & Yung, Y. L. (1990). Two-dimensional atmospheric transport and chemistry model: Numerical experiments with a new advection algorithm. *Journal of Geophysical Research: Atmospheres*, *95*(D6), 7467–7483. https://doi.org/10.1029/JD095iD06p07467

Shia, R.-L., Yung, Y. L., Allen, M., Zurek, R. W., & Crisp, D. (1989). Sensitivity study of advection and diffusion coefficients in a two-dimensional stratospheric model using excess carbon 14 data. *Journal of Geophysical Research: Atmospheres*, *94*(D15), 18467–18484. https://doi.org/10.1029/JD094iD15p18467

Smyshlyaev, S. P., Dvortsov, V. L., Geller, M. A., & Yudin, V. A. (1998). A two-dimensional model with input parameters from a general circulation model: Ozone sensitivity to different formulations for the longitudinal temperature variation. *Journal of Geophysical Research: Atmospheres*, *103*(D21), 28373–28387. https://doi.org/10.1029/98JD02354

Stolzenbach, A., Lefèvre, F., Lebonnois, S., Maattanen, A. E., & Bekki, S. (2015). Three-Dimensional Modelling of Venus Photochemistry. *AGU Fall Meeting Abstracts*, *23*, P23A-2108.

Stolzenbach, A. (2016). *Etude de la photochimie de Vénus à l'aide d'un modèle de circulation générale* [These de doctorat, Paris 6]. https://www.theses.fr/2016PA066413

Taylor, F. W., Beer, R., Chahine, M. T., Diner, D. J., Elson, L. S., Haskins, R. D., McCleese, D. J., Martonchik, J. V., Reichley, P. E., Bradley, S. P., Delderfield, J., Schofield, J. T., Farmer, C. B., Froidevaux, L., Leung, J., Coffey, M. T., & Gille, J. C. (1980). Structure and meteorology of the middle atmosphere of Venus: Infrared remote sensing from the Pioneer Orbiter. *Journal of Geophysical Research: Space Physics*, *85*(A13), 7963–8006. https://doi.org/10.1029/JA085iA13p07963

Vandaele, A. C., Korablev, O., Belyaev, D., Chamberlain, S., Evdokimova, D., Encrenaz, T., Esposito, L., Jessup, K. L., Lefèvre, F., & Limaye, S. (2017a).

Sulfur dioxide in the Venus atmosphere: I. Vertical distribution and variability. *Icarus*, *295*, 16-33.

Vandaele, A. C., Korablev, O., Belyaev, D., Chamberlain, S., Evdokimova, D., Encrenaz, T., Esposito, L., Jessup, K. L., Lefèvre, F., & Limaye, S. (2017b). Sulfur dioxide in the Venus Atmosphere: II. Spatial and temporal variability. *Icarus*, *295*, 1-15.

Vandaele, A. C., Mahieux, A., Chamberlain, S., Ristic, B., Robert, S., Thomas, I. R., Trompet, L., Wilquet, V., & Bertaux, J. L. (2016). Carbon monoxide observed in Venus' atmosphere with SOIR/VEx. *Icarus*, *272*, 48-59. https://doi.org/10.1016/j.icarus.2016.02.025

Yung, Y. L., & DeMore, W. B. (1982). Photochemistry of the stratosphere of Venus: Implications for atmospheric evolution. *Icarus*, *51*(2), 199-247.

Zasova, L. V., Ignatiev, N., Khatuntsev, I., & Linkin, V. (2007). Structure of the Venus atmosphere. *Planetary and Space Science*, *55*(12), 1712-1728. https://doi.org/10.1016/j.pss.2007.01.011

Zasova, L. V., Khatountsev, I. V., Ignatiev, N. I., & Moroz, V. I. (2002). Local time variations of the middle atmosphere of Venus: Solar-related structures. *Advances in Space Research*, *29*(2), 243-248. https://doi.org/10.1016/S0273-1177(01)00574-9

Zhang, X., Liang, M. C., Mills, F. P., Belyaev, D. A., & Yung, Y. L. (2012). Sulfur chemistry in the middle atmosphere of Venus. *Icarus*, *217*(2), 714-739.

Zhang, X., Liang, M.-C., Montmessin, F., Bertaux, J.-L., Parkinson, C., & Yung, Y. L. (2010). Photolysis of sulphuric acid as the source of sulphur oxides in the mesosphere of Venus. *Nature Geoscience*, *3*(12), 834-837.

Zhang, X., & Showman, A. P. (2018). Global-mean Vertical Tracer Mixing in Planetary Atmospheres. I. Theory and Fast-rotating Planets. *The Astrophysical Journal*, *866*(1), 1. https://doi.org/10.3847/1538-4357/aada85